\documentclass[prd,superscriptaddress,twocolumn,nofootinbib,showpacs,preprintnumbers,floatfix]{revtex4}

\usepackage{mathrsfs}
\usepackage{amsfonts,amssymb,amsmath}
\usepackage{graphicx,epsfig}
\usepackage{multirow}
\usepackage{verbatim}

\usepackage{color}
\definecolor{AliceBlue}{rgb}{0.94,0.97,1.00}
\definecolor{AntiqueWhite1}{rgb}{1.00,0.94,0.86}
\definecolor{AntiqueWhite2}{rgb}{0.93,0.87,0.80}
\definecolor{AntiqueWhite3}{rgb}{0.80,0.75,0.69}
\definecolor{AntiqueWhite4}{rgb}{0.55,0.51,0.47}
\definecolor{AntiqueWhite}{rgb}{0.98,0.92,0.84}
\definecolor{BlanchedAlmond}{rgb}{1.00,0.92,0.80}
\definecolor{BlueViolet}{rgb}{0.54,0.17,0.89}
\definecolor{CadetBlue1}{rgb}{0.60,0.96,1.00}
\definecolor{CadetBlue2}{rgb}{0.56,0.90,0.93}
\definecolor{CadetBlue3}{rgb}{0.48,0.77,0.80}
\definecolor{CadetBlue4}{rgb}{0.33,0.53,0.55}
\definecolor{CadetBlue}{rgb}{0.37,0.62,0.63}
\definecolor{CornflowerBlue}{rgb}{0.39,0.58,0.93}
\definecolor{DarkBlue}{rgb}{0.00,0.00,0.55}
\definecolor{DarkCyan}{rgb}{0.00,0.55,0.55}
\definecolor{DarkGoldenrod1}{rgb}{1.00,0.73,0.06}
\definecolor{DarkGoldenrod2}{rgb}{0.93,0.68,0.05}
\definecolor{DarkGoldenrod3}{rgb}{0.80,0.58,0.05}
\definecolor{DarkGoldenrod4}{rgb}{0.55,0.40,0.03}
\definecolor{DarkGoldenrod}{rgb}{0.72,0.53,0.04}
\definecolor{DarkGray}{rgb}{0.66,0.66,0.66}
\definecolor{DarkGreen}{rgb}{0.00,0.39,0.00}
\definecolor{DarkGrey}{rgb}{0.66,0.66,0.66}
\definecolor{DarkKhaki}{rgb}{0.74,0.72,0.42}
\definecolor{DarkMagenta}{rgb}{0.55,0.00,0.55}
\definecolor{DarkOliveGreen1}{rgb}{0.79,1.00,0.44}
\definecolor{DarkOliveGreen2}{rgb}{0.74,0.93,0.41}
\definecolor{DarkOliveGreen3}{rgb}{0.64,0.80,0.35}
\definecolor{DarkOliveGreen4}{rgb}{0.43,0.55,0.24}
\definecolor{DarkOliveGreen}{rgb}{0.33,0.42,0.18}
\definecolor{DarkOrange1}{rgb}{1.00,0.50,0.00}
\definecolor{DarkOrange2}{rgb}{0.93,0.46,0.00}
\definecolor{DarkOrange3}{rgb}{0.80,0.40,0.00}
\definecolor{DarkOrange4}{rgb}{0.55,0.27,0.00}
\definecolor{DarkOrange}{rgb}{1.00,0.55,0.00}
\definecolor{DarkOrchid1}{rgb}{0.75,0.24,1.00}
\definecolor{DarkOrchid2}{rgb}{0.70,0.23,0.93}
\definecolor{DarkOrchid3}{rgb}{0.60,0.20,0.80}
\definecolor{DarkOrchid4}{rgb}{0.41,0.13,0.55}
\definecolor{DarkOrchid}{rgb}{0.60,0.20,0.80}
\definecolor{DarkRed}{rgb}{0.55,0.00,0.00}
\definecolor{DarkSalmon}{rgb}{0.91,0.59,0.48}
\definecolor{DarkSeaGreen1}{rgb}{0.76,1.00,0.76}
\definecolor{DarkSeaGreen2}{rgb}{0.71,0.93,0.71}
\definecolor{DarkSeaGreen3}{rgb}{0.61,0.80,0.61}
\definecolor{DarkSeaGreen4}{rgb}{0.41,0.55,0.41}
\definecolor{DarkSeaGreen}{rgb}{0.56,0.74,0.56}
\definecolor{DarkSlateBlue}{rgb}{0.28,0.24,0.55}
\definecolor{DarkSlateGray1}{rgb}{0.59,1.00,1.00}
\definecolor{DarkSlateGray2}{rgb}{0.55,0.93,0.93}
\definecolor{DarkSlateGray3}{rgb}{0.47,0.80,0.80}
\definecolor{DarkSlateGray4}{rgb}{0.32,0.55,0.55}
\definecolor{DarkSlateGray}{rgb}{0.18,0.31,0.31}
\definecolor{DarkSlateGrey}{rgb}{0.18,0.31,0.31}
\definecolor{DarkTurquoise}{rgb}{0.00,0.81,0.82}
\definecolor{DarkViolet}{rgb}{0.58,0.00,0.83}
\definecolor{DeepPink1}{rgb}{1.00,0.08,0.58}
\definecolor{DeepPink2}{rgb}{0.93,0.07,0.54}
\definecolor{DeepPink3}{rgb}{0.80,0.06,0.46}
\definecolor{DeepPink4}{rgb}{0.55,0.04,0.31}
\definecolor{DeepPink}{rgb}{1.00,0.08,0.58}
\definecolor{DeepSkyBlue1}{rgb}{0.00,0.75,1.00}
\definecolor{DeepSkyBlue2}{rgb}{0.00,0.70,0.93}
\definecolor{DeepSkyBlue3}{rgb}{0.00,0.60,0.80}
\definecolor{DeepSkyBlue4}{rgb}{0.00,0.41,0.55}
\definecolor{DeepSkyBlue}{rgb}{0.00,0.75,1.00}
\definecolor{DimGray}{rgb}{0.41,0.41,0.41}
\definecolor{DimGrey}{rgb}{0.41,0.41,0.41}
\definecolor{DodgerBlue1}{rgb}{0.12,0.56,1.00}
\definecolor{DodgerBlue2}{rgb}{0.11,0.53,0.93}
\definecolor{DodgerBlue3}{rgb}{0.09,0.45,0.80}
\definecolor{DodgerBlue4}{rgb}{0.06,0.31,0.55}
\definecolor{DodgerBlue}{rgb}{0.12,0.56,1.00}
\definecolor{FloralWhite}{rgb}{1.00,0.98,0.94}
\definecolor{ForestGreen}{rgb}{0.13,0.55,0.13}
\definecolor{GhostWhite}{rgb}{0.97,0.97,1.00}
\definecolor{GreenYellow}{rgb}{0.68,1.00,0.18}
\definecolor{HotPink1}{rgb}{1.00,0.43,0.71}
\definecolor{HotPink2}{rgb}{0.93,0.42,0.65}
\definecolor{HotPink3}{rgb}{0.80,0.38,0.56}
\definecolor{HotPink4}{rgb}{0.55,0.23,0.38}
\definecolor{HotPink}{rgb}{1.00,0.41,0.71}
\definecolor{IndianRed1}{rgb}{1.00,0.42,0.42}
\definecolor{IndianRed2}{rgb}{0.93,0.39,0.39}
\definecolor{IndianRed3}{rgb}{0.80,0.33,0.33}
\definecolor{IndianRed4}{rgb}{0.55,0.23,0.23}
\definecolor{IndianRed}{rgb}{0.80,0.36,0.36}
\definecolor{LavenderBlush1}{rgb}{1.00,0.94,0.96}
\definecolor{LavenderBlush2}{rgb}{0.93,0.88,0.90}
\definecolor{LavenderBlush3}{rgb}{0.80,0.76,0.77}
\definecolor{LavenderBlush4}{rgb}{0.55,0.51,0.53}
\definecolor{LavenderBlush}{rgb}{1.00,0.94,0.96}
\definecolor{LawnGreen}{rgb}{0.49,0.99,0.00}
\definecolor{LemonChiffon1}{rgb}{1.00,0.98,0.80}
\definecolor{LemonChiffon2}{rgb}{0.93,0.91,0.75}
\definecolor{LemonChiffon3}{rgb}{0.80,0.79,0.65}
\definecolor{LemonChiffon4}{rgb}{0.55,0.54,0.44}
\definecolor{LemonChiffon}{rgb}{1.00,0.98,0.80}
\definecolor{LightBlue1}{rgb}{0.75,0.94,1.00}
\definecolor{LightBlue2}{rgb}{0.70,0.87,0.93}
\definecolor{LightBlue3}{rgb}{0.60,0.75,0.80}
\definecolor{LightBlue4}{rgb}{0.41,0.51,0.55}
\definecolor{LightBlue}{rgb}{0.68,0.85,0.90}
\definecolor{LightCoral}{rgb}{0.94,0.50,0.50}
\definecolor{LightCyan1}{rgb}{0.88,1.00,1.00}
\definecolor{LightCyan2}{rgb}{0.82,0.93,0.93}
\definecolor{LightCyan3}{rgb}{0.71,0.80,0.80}
\definecolor{LightCyan4}{rgb}{0.48,0.55,0.55}
\definecolor{LightCyan}{rgb}{0.88,1.00,1.00}
\definecolor{LightGoldenrod1}{rgb}{1.00,0.93,0.55}
\definecolor{LightGoldenrod2}{rgb}{0.93,0.86,0.51}
\definecolor{LightGoldenrod3}{rgb}{0.80,0.75,0.44}
\definecolor{LightGoldenrod4}{rgb}{0.55,0.51,0.30}
\definecolor{LightGoldenrodYellow}{rgb}{0.98,0.98,0.82}
\definecolor{LightGoldenrod}{rgb}{0.93,0.87,0.51}
\definecolor{LightGray}{rgb}{0.83,0.83,0.83}
\definecolor{LightGreen}{rgb}{0.56,0.93,0.56}
\definecolor{LightGrey}{rgb}{0.83,0.83,0.83}
\definecolor{LightPink1}{rgb}{1.00,0.68,0.73}
\definecolor{LightPink2}{rgb}{0.93,0.64,0.68}
\definecolor{LightPink3}{rgb}{0.80,0.55,0.58}
\definecolor{LightPink4}{rgb}{0.55,0.37,0.40}
\definecolor{LightPink}{rgb}{1.00,0.71,0.76}
\definecolor{LightSalmon1}{rgb}{1.00,0.63,0.48}
\definecolor{LightSalmon2}{rgb}{0.93,0.58,0.45}
\definecolor{LightSalmon3}{rgb}{0.80,0.51,0.38}
\definecolor{LightSalmon4}{rgb}{0.55,0.34,0.26}
\definecolor{LightSalmon}{rgb}{1.00,0.63,0.48}
\definecolor{LightSeaGreen}{rgb}{0.13,0.70,0.67}
\definecolor{LightSkyBlue1}{rgb}{0.69,0.89,1.00}
\definecolor{LightSkyBlue2}{rgb}{0.64,0.83,0.93}
\definecolor{LightSkyBlue3}{rgb}{0.55,0.71,0.80}
\definecolor{LightSkyBlue4}{rgb}{0.38,0.48,0.55}
\definecolor{LightSkyBlue}{rgb}{0.53,0.81,0.98}
\definecolor{LightSlateBlue}{rgb}{0.52,0.44,1.00}
\definecolor{LightSlateGray}{rgb}{0.47,0.53,0.60}
\definecolor{LightSlateGrey}{rgb}{0.47,0.53,0.60}
\definecolor{LightSteelBlue1}{rgb}{0.79,0.88,1.00}
\definecolor{LightSteelBlue2}{rgb}{0.74,0.82,0.93}
\definecolor{LightSteelBlue3}{rgb}{0.64,0.71,0.80}
\definecolor{LightSteelBlue4}{rgb}{0.43,0.48,0.55}
\definecolor{LightSteelBlue}{rgb}{0.69,0.77,0.87}
\definecolor{LightYellow1}{rgb}{1.00,1.00,0.88}
\definecolor{LightYellow2}{rgb}{0.93,0.93,0.82}
\definecolor{LightYellow3}{rgb}{0.80,0.80,0.71}
\definecolor{LightYellow4}{rgb}{0.55,0.55,0.48}
\definecolor{LightYellow}{rgb}{1.00,1.00,0.88}
\definecolor{LimeGreen}{rgb}{0.20,0.80,0.20}
\definecolor{MediumAquamarine}{rgb}{0.40,0.80,0.67}
\definecolor{MediumBlue}{rgb}{0.00,0.00,0.80}
\definecolor{MediumOrchid1}{rgb}{0.88,0.40,1.00}
\definecolor{MediumOrchid2}{rgb}{0.82,0.37,0.93}
\definecolor{MediumOrchid3}{rgb}{0.71,0.32,0.80}
\definecolor{MediumOrchid4}{rgb}{0.48,0.22,0.55}
\definecolor{MediumOrchid}{rgb}{0.73,0.33,0.83}
\definecolor{MediumPurple1}{rgb}{0.67,0.51,1.00}
\definecolor{MediumPurple2}{rgb}{0.62,0.47,0.93}
\definecolor{MediumPurple3}{rgb}{0.54,0.41,0.80}
\definecolor{MediumPurple4}{rgb}{0.36,0.28,0.55}
\definecolor{MediumPurple}{rgb}{0.58,0.44,0.86}
\definecolor{MediumSeaGreen}{rgb}{0.24,0.70,0.44}
\definecolor{MediumSlateBlue}{rgb}{0.48,0.41,0.93}
\definecolor{MediumSpringGreen}{rgb}{0.00,0.98,0.60}
\definecolor{MediumTurquoise}{rgb}{0.28,0.82,0.80}
\definecolor{MediumVioletRed}{rgb}{0.78,0.08,0.52}
\definecolor{MidnightBlue}{rgb}{0.10,0.10,0.44}
\definecolor{MintCream}{rgb}{0.96,1.00,0.98}
\definecolor{MistyRose1}{rgb}{1.00,0.89,0.88}
\definecolor{MistyRose2}{rgb}{0.93,0.84,0.82}
\definecolor{MistyRose3}{rgb}{0.80,0.72,0.71}
\definecolor{MistyRose4}{rgb}{0.55,0.49,0.48}
\definecolor{MistyRose}{rgb}{1.00,0.89,0.88}
\definecolor{NavajoWhite1}{rgb}{1.00,0.87,0.68}
\definecolor{NavajoWhite2}{rgb}{0.93,0.81,0.63}
\definecolor{NavajoWhite3}{rgb}{0.80,0.70,0.55}
\definecolor{NavajoWhite4}{rgb}{0.55,0.47,0.37}
\definecolor{NavajoWhite}{rgb}{1.00,0.87,0.68}
\definecolor{NavyBlue}{rgb}{0.00,0.00,0.50}
\definecolor{OldLace}{rgb}{0.99,0.96,0.90}
\definecolor{OliveDrab1}{rgb}{0.75,1.00,0.24}
\definecolor{OliveDrab2}{rgb}{0.70,0.93,0.23}
\definecolor{OliveDrab3}{rgb}{0.60,0.80,0.20}
\definecolor{OliveDrab4}{rgb}{0.41,0.55,0.13}
\definecolor{OliveDrab}{rgb}{0.42,0.56,0.14}
\definecolor{OrangeRed1}{rgb}{1.00,0.27,0.00}
\definecolor{OrangeRed2}{rgb}{0.93,0.25,0.00}
\definecolor{OrangeRed3}{rgb}{0.80,0.22,0.00}
\definecolor{OrangeRed4}{rgb}{0.55,0.15,0.00}
\definecolor{OrangeRed}{rgb}{1.00,0.27,0.00}
\definecolor{PaleGoldenrod}{rgb}{0.93,0.91,0.67}
\definecolor{PaleGreen1}{rgb}{0.60,1.00,0.60}
\definecolor{PaleGreen2}{rgb}{0.56,0.93,0.56}
\definecolor{PaleGreen3}{rgb}{0.49,0.80,0.49}
\definecolor{PaleGreen4}{rgb}{0.33,0.55,0.33}
\definecolor{PaleGreen}{rgb}{0.60,0.98,0.60}
\definecolor{PaleTurquoise1}{rgb}{0.73,1.00,1.00}
\definecolor{PaleTurquoise2}{rgb}{0.68,0.93,0.93}
\definecolor{PaleTurquoise3}{rgb}{0.59,0.80,0.80}
\definecolor{PaleTurquoise4}{rgb}{0.40,0.55,0.55}
\definecolor{PaleTurquoise}{rgb}{0.69,0.93,0.93}
\definecolor{PaleVioletRed1}{rgb}{1.00,0.51,0.67}
\definecolor{PaleVioletRed2}{rgb}{0.93,0.47,0.62}
\definecolor{PaleVioletRed3}{rgb}{0.80,0.41,0.54}
\definecolor{PaleVioletRed4}{rgb}{0.55,0.28,0.36}
\definecolor{PaleVioletRed}{rgb}{0.86,0.44,0.58}
\definecolor{PapayaWhip}{rgb}{1.00,0.94,0.84}
\definecolor{PeachPuff1}{rgb}{1.00,0.85,0.73}
\definecolor{PeachPuff2}{rgb}{0.93,0.80,0.68}
\definecolor{PeachPuff3}{rgb}{0.80,0.69,0.58}
\definecolor{PeachPuff4}{rgb}{0.55,0.47,0.40}
\definecolor{PeachPuff}{rgb}{1.00,0.85,0.73}
\definecolor{PowderBlue}{rgb}{0.69,0.88,0.90}
\definecolor{RosyBrown1}{rgb}{1.00,0.76,0.76}
\definecolor{RosyBrown2}{rgb}{0.93,0.71,0.71}
\definecolor{RosyBrown3}{rgb}{0.80,0.61,0.61}
\definecolor{RosyBrown4}{rgb}{0.55,0.41,0.41}
\definecolor{RosyBrown}{rgb}{0.74,0.56,0.56}
\definecolor{RoyalBlue1}{rgb}{0.28,0.46,1.00}
\definecolor{RoyalBlue2}{rgb}{0.26,0.43,0.93}
\definecolor{RoyalBlue3}{rgb}{0.23,0.37,0.80}
\definecolor{RoyalBlue4}{rgb}{0.15,0.25,0.55}
\definecolor{RoyalBlue}{rgb}{0.25,0.41,0.88}
\definecolor{SaddleBrown}{rgb}{0.55,0.27,0.07}
\definecolor{SandyBrown}{rgb}{0.96,0.64,0.38}
\definecolor{SeaGreen1}{rgb}{0.33,1.00,0.62}
\definecolor{SeaGreen2}{rgb}{0.31,0.93,0.58}
\definecolor{SeaGreen3}{rgb}{0.26,0.80,0.50}
\definecolor{SeaGreen4}{rgb}{0.18,0.55,0.34}
\definecolor{SeaGreen}{rgb}{0.18,0.55,0.34}
\definecolor{SkyBlue1}{rgb}{0.53,0.81,1.00}
\definecolor{SkyBlue2}{rgb}{0.49,0.75,0.93}
\definecolor{SkyBlue3}{rgb}{0.42,0.65,0.80}
\definecolor{SkyBlue4}{rgb}{0.29,0.44,0.55}
\definecolor{SkyBlue}{rgb}{0.53,0.81,0.92}
\definecolor{SlateBlue1}{rgb}{0.51,0.44,1.00}
\definecolor{SlateBlue2}{rgb}{0.48,0.40,0.93}
\definecolor{SlateBlue3}{rgb}{0.41,0.35,0.80}
\definecolor{SlateBlue4}{rgb}{0.28,0.24,0.55}
\definecolor{SlateBlue}{rgb}{0.42,0.35,0.80}
\definecolor{SlateGray1}{rgb}{0.78,0.89,1.00}
\definecolor{SlateGray2}{rgb}{0.73,0.83,0.93}
\definecolor{SlateGray3}{rgb}{0.62,0.71,0.80}
\definecolor{SlateGray4}{rgb}{0.42,0.48,0.55}
\definecolor{SlateGray}{rgb}{0.44,0.50,0.56}
\definecolor{SlateGrey}{rgb}{0.44,0.50,0.56}
\definecolor{SpringGreen1}{rgb}{0.00,1.00,0.50}
\definecolor{SpringGreen2}{rgb}{0.00,0.93,0.46}
\definecolor{SpringGreen3}{rgb}{0.00,0.80,0.40}
\definecolor{SpringGreen4}{rgb}{0.00,0.55,0.27}
\definecolor{SpringGreen}{rgb}{0.00,1.00,0.50}
\definecolor{SteelBlue1}{rgb}{0.39,0.72,1.00}
\definecolor{SteelBlue2}{rgb}{0.36,0.67,0.93}
\definecolor{SteelBlue3}{rgb}{0.31,0.58,0.80}
\definecolor{SteelBlue4}{rgb}{0.21,0.39,0.55}
\definecolor{SteelBlue}{rgb}{0.27,0.51,0.71}
\definecolor{VioletRed1}{rgb}{1.00,0.24,0.59}
\definecolor{VioletRed2}{rgb}{0.93,0.23,0.55}
\definecolor{VioletRed3}{rgb}{0.80,0.20,0.47}
\definecolor{VioletRed4}{rgb}{0.55,0.13,0.32}
\definecolor{VioletRed}{rgb}{0.82,0.13,0.56}
\definecolor{WhiteSmoke}{rgb}{0.96,0.96,0.96}
\definecolor{YellowGreen}{rgb}{0.60,0.80,0.20}
\definecolor{aliceblue}{rgb}{0.94,0.97,1.00}
\definecolor{antiquewhite}{rgb}{0.98,0.92,0.84}
\definecolor{aquamarine1}{rgb}{0.50,1.00,0.83}
\definecolor{aquamarine2}{rgb}{0.46,0.93,0.78}
\definecolor{aquamarine3}{rgb}{0.40,0.80,0.67}
\definecolor{aquamarine4}{rgb}{0.27,0.55,0.45}
\definecolor{aquamarine}{rgb}{0.50,1.00,0.83}
\definecolor{azure1}{rgb}{0.94,1.00,1.00}
\definecolor{azure2}{rgb}{0.88,0.93,0.93}
\definecolor{azure3}{rgb}{0.76,0.80,0.80}
\definecolor{azure4}{rgb}{0.51,0.55,0.55}
\definecolor{azure}{rgb}{0.94,1.00,1.00}
\definecolor{beige}{rgb}{0.96,0.96,0.86}
\definecolor{bisque1}{rgb}{1.00,0.89,0.77}
\definecolor{bisque2}{rgb}{0.93,0.84,0.72}
\definecolor{bisque3}{rgb}{0.80,0.72,0.62}
\definecolor{bisque4}{rgb}{0.55,0.49,0.42}
\definecolor{bisque}{rgb}{1.00,0.89,0.77}
\definecolor{black}{rgb}{0.00,0.00,0.00}
\definecolor{blanchedalmond}{rgb}{1.00,0.92,0.80}
\definecolor{blue1}{rgb}{0.00,0.00,1.00}
\definecolor{blue2}{rgb}{0.00,0.00,0.93}
\definecolor{blue3}{rgb}{0.00,0.00,0.80}
\definecolor{blue4}{rgb}{0.00,0.00,0.55}
\definecolor{blueviolet}{rgb}{0.54,0.17,0.89}
\definecolor{blue}{rgb}{0.00,0.00,1.00}
\definecolor{brown1}{rgb}{1.00,0.25,0.25}
\definecolor{brown2}{rgb}{0.93,0.23,0.23}
\definecolor{brown3}{rgb}{0.80,0.20,0.20}
\definecolor{brown4}{rgb}{0.55,0.14,0.14}
\definecolor{brown}{rgb}{0.65,0.16,0.16}
\definecolor{burlywood1}{rgb}{1.00,0.83,0.61}
\definecolor{burlywood2}{rgb}{0.93,0.77,0.57}
\definecolor{burlywood3}{rgb}{0.80,0.67,0.49}
\definecolor{burlywood4}{rgb}{0.55,0.45,0.33}
\definecolor{burlywood}{rgb}{0.87,0.72,0.53}
\definecolor{cadetblue}{rgb}{0.37,0.62,0.63}
\definecolor{chartreuse1}{rgb}{0.50,1.00,0.00}
\definecolor{chartreuse2}{rgb}{0.46,0.93,0.00}
\definecolor{chartreuse3}{rgb}{0.40,0.80,0.00}
\definecolor{chartreuse4}{rgb}{0.27,0.55,0.00}
\definecolor{chartreuse}{rgb}{0.50,1.00,0.00}
\definecolor{chocolate1}{rgb}{1.00,0.50,0.14}
\definecolor{chocolate2}{rgb}{0.93,0.46,0.13}
\definecolor{chocolate3}{rgb}{0.80,0.40,0.11}
\definecolor{chocolate4}{rgb}{0.55,0.27,0.07}
\definecolor{chocolate}{rgb}{0.82,0.41,0.12}
\definecolor{coral1}{rgb}{1.00,0.45,0.34}
\definecolor{coral2}{rgb}{0.93,0.42,0.31}
\definecolor{coral3}{rgb}{0.80,0.36,0.27}
\definecolor{coral4}{rgb}{0.55,0.24,0.18}
\definecolor{coral}{rgb}{1.00,0.50,0.31}
\definecolor{cornflowerblue}{rgb}{0.39,0.58,0.93}
\definecolor{cornsilk1}{rgb}{1.00,0.97,0.86}
\definecolor{cornsilk2}{rgb}{0.93,0.91,0.80}
\definecolor{cornsilk3}{rgb}{0.80,0.78,0.69}
\definecolor{cornsilk4}{rgb}{0.55,0.53,0.47}
\definecolor{cornsilk}{rgb}{1.00,0.97,0.86}
\definecolor{cyan1}{rgb}{0.00,1.00,1.00}
\definecolor{cyan2}{rgb}{0.00,0.93,0.93}
\definecolor{cyan3}{rgb}{0.00,0.80,0.80}
\definecolor{cyan4}{rgb}{0.00,0.55,0.55}
\definecolor{cyan}{rgb}{0.00,1.00,1.00}
\definecolor{darkblue}{rgb}{0.00,0.00,0.55}
\definecolor{darkcyan}{rgb}{0.00,0.55,0.55}
\definecolor{darkgoldenrod}{rgb}{0.72,0.53,0.04}
\definecolor{darkgray}{rgb}{0.66,0.66,0.66}
\definecolor{darkgreen}{rgb}{0.00,0.39,0.00}
\definecolor{darkgrey}{rgb}{0.66,0.66,0.66}
\definecolor{darkkhaki}{rgb}{0.74,0.72,0.42}
\definecolor{darkmagenta}{rgb}{0.55,0.00,0.55}
\definecolor{darkolive}{rgb}{0.33,0.42,0.18}
\definecolor{darkorange}{rgb}{1.00,0.55,0.00}
\definecolor{darkorchid}{rgb}{0.60,0.20,0.80}
\definecolor{darkred}{rgb}{0.55,0.00,0.00}
\definecolor{darksalmon}{rgb}{0.91,0.59,0.48}
\definecolor{darksea}{rgb}{0.56,0.74,0.56}
\definecolor{darkslate}{rgb}{0.18,0.31,0.31}
\definecolor{darkslate}{rgb}{0.18,0.31,0.31}
\definecolor{darkslate}{rgb}{0.28,0.24,0.55}
\definecolor{darkturquoise}{rgb}{0.00,0.81,0.82}
\definecolor{darkviolet}{rgb}{0.58,0.00,0.83}
\definecolor{deeppink}{rgb}{1.00,0.08,0.58}
\definecolor{deepsky}{rgb}{0.00,0.75,1.00}
\definecolor{dimgray}{rgb}{0.41,0.41,0.41}
\definecolor{dimgrey}{rgb}{0.41,0.41,0.41}
\definecolor{dodgerblue}{rgb}{0.12,0.56,1.00}
\definecolor{firebrick1}{rgb}{1.00,0.19,0.19}
\definecolor{firebrick2}{rgb}{0.93,0.17,0.17}
\definecolor{firebrick3}{rgb}{0.80,0.15,0.15}
\definecolor{firebrick4}{rgb}{0.55,0.10,0.10}
\definecolor{firebrick}{rgb}{0.70,0.13,0.13}
\definecolor{floralwhite}{rgb}{1.00,0.98,0.94}
\definecolor{forestgreen}{rgb}{0.13,0.55,0.13}
\definecolor{gainsboro}{rgb}{0.86,0.86,0.86}
\definecolor{ghostwhite}{rgb}{0.97,0.97,1.00}
\definecolor{gold1}{rgb}{1.00,0.84,0.00}
\definecolor{gold2}{rgb}{0.93,0.79,0.00}
\definecolor{gold3}{rgb}{0.80,0.68,0.00}
\definecolor{gold4}{rgb}{0.55,0.46,0.00}
\definecolor{goldenrod1}{rgb}{1.00,0.76,0.15}
\definecolor{goldenrod2}{rgb}{0.93,0.71,0.13}
\definecolor{goldenrod3}{rgb}{0.80,0.61,0.11}
\definecolor{goldenrod4}{rgb}{0.55,0.41,0.08}
\definecolor{goldenrod}{rgb}{0.85,0.65,0.13}
\definecolor{gold}{rgb}{1.00,0.84,0.00}
\definecolor{gray0}{rgb}{0.00,0.00,0.00}
\definecolor{gray100}{rgb}{1.00,1.00,1.00}
\definecolor{gray10}{rgb}{0.10,0.10,0.10}
\definecolor{gray11}{rgb}{0.11,0.11,0.11}
\definecolor{gray12}{rgb}{0.12,0.12,0.12}
\definecolor{gray13}{rgb}{0.13,0.13,0.13}
\definecolor{gray14}{rgb}{0.14,0.14,0.14}
\definecolor{gray15}{rgb}{0.15,0.15,0.15}
\definecolor{gray16}{rgb}{0.16,0.16,0.16}
\definecolor{gray17}{rgb}{0.17,0.17,0.17}
\definecolor{gray18}{rgb}{0.18,0.18,0.18}
\definecolor{gray19}{rgb}{0.19,0.19,0.19}
\definecolor{gray1}{rgb}{0.01,0.01,0.01}
\definecolor{gray20}{rgb}{0.20,0.20,0.20}
\definecolor{gray21}{rgb}{0.21,0.21,0.21}
\definecolor{gray22}{rgb}{0.22,0.22,0.22}
\definecolor{gray23}{rgb}{0.23,0.23,0.23}
\definecolor{gray24}{rgb}{0.24,0.24,0.24}
\definecolor{gray25}{rgb}{0.25,0.25,0.25}
\definecolor{gray26}{rgb}{0.26,0.26,0.26}
\definecolor{gray27}{rgb}{0.27,0.27,0.27}
\definecolor{gray28}{rgb}{0.28,0.28,0.28}
\definecolor{gray29}{rgb}{0.29,0.29,0.29}
\definecolor{gray2}{rgb}{0.02,0.02,0.02}
\definecolor{gray30}{rgb}{0.30,0.30,0.30}
\definecolor{gray31}{rgb}{0.31,0.31,0.31}
\definecolor{gray32}{rgb}{0.32,0.32,0.32}
\definecolor{gray33}{rgb}{0.33,0.33,0.33}
\definecolor{gray34}{rgb}{0.34,0.34,0.34}
\definecolor{gray35}{rgb}{0.35,0.35,0.35}
\definecolor{gray36}{rgb}{0.36,0.36,0.36}
\definecolor{gray37}{rgb}{0.37,0.37,0.37}
\definecolor{gray38}{rgb}{0.38,0.38,0.38}
\definecolor{gray39}{rgb}{0.39,0.39,0.39}
\definecolor{gray3}{rgb}{0.03,0.03,0.03}
\definecolor{gray40}{rgb}{0.40,0.40,0.40}
\definecolor{gray41}{rgb}{0.41,0.41,0.41}
\definecolor{gray42}{rgb}{0.42,0.42,0.42}
\definecolor{gray43}{rgb}{0.43,0.43,0.43}
\definecolor{gray44}{rgb}{0.44,0.44,0.44}
\definecolor{gray45}{rgb}{0.45,0.45,0.45}
\definecolor{gray46}{rgb}{0.46,0.46,0.46}
\definecolor{gray47}{rgb}{0.47,0.47,0.47}
\definecolor{gray48}{rgb}{0.48,0.48,0.48}
\definecolor{gray49}{rgb}{0.49,0.49,0.49}
\definecolor{gray4}{rgb}{0.04,0.04,0.04}
\definecolor{gray50}{rgb}{0.50,0.50,0.50}
\definecolor{gray51}{rgb}{0.51,0.51,0.51}
\definecolor{gray52}{rgb}{0.52,0.52,0.52}
\definecolor{gray53}{rgb}{0.53,0.53,0.53}
\definecolor{gray54}{rgb}{0.54,0.54,0.54}
\definecolor{gray55}{rgb}{0.55,0.55,0.55}
\definecolor{gray56}{rgb}{0.56,0.56,0.56}
\definecolor{gray57}{rgb}{0.57,0.57,0.57}
\definecolor{gray58}{rgb}{0.58,0.58,0.58}
\definecolor{gray59}{rgb}{0.59,0.59,0.59}
\definecolor{gray5}{rgb}{0.05,0.05,0.05}
\definecolor{gray60}{rgb}{0.60,0.60,0.60}
\definecolor{gray61}{rgb}{0.61,0.61,0.61}
\definecolor{gray62}{rgb}{0.62,0.62,0.62}
\definecolor{gray63}{rgb}{0.63,0.63,0.63}
\definecolor{gray64}{rgb}{0.64,0.64,0.64}
\definecolor{gray65}{rgb}{0.65,0.65,0.65}
\definecolor{gray66}{rgb}{0.66,0.66,0.66}
\definecolor{gray67}{rgb}{0.67,0.67,0.67}
\definecolor{gray68}{rgb}{0.68,0.68,0.68}
\definecolor{gray69}{rgb}{0.69,0.69,0.69}
\definecolor{gray6}{rgb}{0.06,0.06,0.06}
\definecolor{gray70}{rgb}{0.70,0.70,0.70}
\definecolor{gray71}{rgb}{0.71,0.71,0.71}
\definecolor{gray72}{rgb}{0.72,0.72,0.72}
\definecolor{gray73}{rgb}{0.73,0.73,0.73}
\definecolor{gray74}{rgb}{0.74,0.74,0.74}
\definecolor{gray75}{rgb}{0.75,0.75,0.75}
\definecolor{gray76}{rgb}{0.76,0.76,0.76}
\definecolor{gray77}{rgb}{0.77,0.77,0.77}
\definecolor{gray78}{rgb}{0.78,0.78,0.78}
\definecolor{gray79}{rgb}{0.79,0.79,0.79}
\definecolor{gray7}{rgb}{0.07,0.07,0.07}
\definecolor{gray80}{rgb}{0.80,0.80,0.80}
\definecolor{gray81}{rgb}{0.81,0.81,0.81}
\definecolor{gray82}{rgb}{0.82,0.82,0.82}
\definecolor{gray83}{rgb}{0.83,0.83,0.83}
\definecolor{gray84}{rgb}{0.84,0.84,0.84}
\definecolor{gray85}{rgb}{0.85,0.85,0.85}
\definecolor{gray86}{rgb}{0.86,0.86,0.86}
\definecolor{gray87}{rgb}{0.87,0.87,0.87}
\definecolor{gray88}{rgb}{0.88,0.88,0.88}
\definecolor{gray89}{rgb}{0.89,0.89,0.89}
\definecolor{gray8}{rgb}{0.08,0.08,0.08}
\definecolor{gray90}{rgb}{0.90,0.90,0.90}
\definecolor{gray91}{rgb}{0.91,0.91,0.91}
\definecolor{gray92}{rgb}{0.92,0.92,0.92}
\definecolor{gray93}{rgb}{0.93,0.93,0.93}
\definecolor{gray94}{rgb}{0.94,0.94,0.94}
\definecolor{gray95}{rgb}{0.95,0.95,0.95}
\definecolor{gray96}{rgb}{0.96,0.96,0.96}
\definecolor{gray97}{rgb}{0.97,0.97,0.97}
\definecolor{gray98}{rgb}{0.98,0.98,0.98}
\definecolor{gray99}{rgb}{0.99,0.99,0.99}
\definecolor{gray9}{rgb}{0.09,0.09,0.09}
\definecolor{gray}{rgb}{0.75,0.75,0.75}
\definecolor{green1}{rgb}{0.00,1.00,0.00}
\definecolor{green2}{rgb}{0.00,0.93,0.00}
\definecolor{green3}{rgb}{0.00,0.80,0.00}
\definecolor{green4}{rgb}{0.00,0.55,0.00}
\definecolor{greenyellow}{rgb}{0.68,1.00,0.18}
\definecolor{green}{rgb}{0.00,1.00,0.00}
\definecolor{grey0}{rgb}{0.00,0.00,0.00}
\definecolor{grey100}{rgb}{1.00,1.00,1.00}
\definecolor{grey10}{rgb}{0.10,0.10,0.10}
\definecolor{grey11}{rgb}{0.11,0.11,0.11}
\definecolor{grey12}{rgb}{0.12,0.12,0.12}
\definecolor{grey13}{rgb}{0.13,0.13,0.13}
\definecolor{grey14}{rgb}{0.14,0.14,0.14}
\definecolor{grey15}{rgb}{0.15,0.15,0.15}
\definecolor{grey16}{rgb}{0.16,0.16,0.16}
\definecolor{grey17}{rgb}{0.17,0.17,0.17}
\definecolor{grey18}{rgb}{0.18,0.18,0.18}
\definecolor{grey19}{rgb}{0.19,0.19,0.19}
\definecolor{grey1}{rgb}{0.01,0.01,0.01}
\definecolor{grey20}{rgb}{0.20,0.20,0.20}
\definecolor{grey21}{rgb}{0.21,0.21,0.21}
\definecolor{grey22}{rgb}{0.22,0.22,0.22}
\definecolor{grey23}{rgb}{0.23,0.23,0.23}
\definecolor{grey24}{rgb}{0.24,0.24,0.24}
\definecolor{grey25}{rgb}{0.25,0.25,0.25}
\definecolor{grey26}{rgb}{0.26,0.26,0.26}
\definecolor{grey27}{rgb}{0.27,0.27,0.27}
\definecolor{grey28}{rgb}{0.28,0.28,0.28}
\definecolor{grey29}{rgb}{0.29,0.29,0.29}
\definecolor{grey2}{rgb}{0.02,0.02,0.02}
\definecolor{grey30}{rgb}{0.30,0.30,0.30}
\definecolor{grey31}{rgb}{0.31,0.31,0.31}
\definecolor{grey32}{rgb}{0.32,0.32,0.32}
\definecolor{grey33}{rgb}{0.33,0.33,0.33}
\definecolor{grey34}{rgb}{0.34,0.34,0.34}
\definecolor{grey35}{rgb}{0.35,0.35,0.35}
\definecolor{grey36}{rgb}{0.36,0.36,0.36}
\definecolor{grey37}{rgb}{0.37,0.37,0.37}
\definecolor{grey38}{rgb}{0.38,0.38,0.38}
\definecolor{grey39}{rgb}{0.39,0.39,0.39}
\definecolor{grey3}{rgb}{0.03,0.03,0.03}
\definecolor{grey40}{rgb}{0.40,0.40,0.40}
\definecolor{grey41}{rgb}{0.41,0.41,0.41}
\definecolor{grey42}{rgb}{0.42,0.42,0.42}
\definecolor{grey43}{rgb}{0.43,0.43,0.43}
\definecolor{grey44}{rgb}{0.44,0.44,0.44}
\definecolor{grey45}{rgb}{0.45,0.45,0.45}
\definecolor{grey46}{rgb}{0.46,0.46,0.46}
\definecolor{grey47}{rgb}{0.47,0.47,0.47}
\definecolor{grey48}{rgb}{0.48,0.48,0.48}
\definecolor{grey49}{rgb}{0.49,0.49,0.49}
\definecolor{grey4}{rgb}{0.04,0.04,0.04}
\definecolor{grey50}{rgb}{0.50,0.50,0.50}
\definecolor{grey51}{rgb}{0.51,0.51,0.51}
\definecolor{grey52}{rgb}{0.52,0.52,0.52}
\definecolor{grey53}{rgb}{0.53,0.53,0.53}
\definecolor{grey54}{rgb}{0.54,0.54,0.54}
\definecolor{grey55}{rgb}{0.55,0.55,0.55}
\definecolor{grey56}{rgb}{0.56,0.56,0.56}
\definecolor{grey57}{rgb}{0.57,0.57,0.57}
\definecolor{grey58}{rgb}{0.58,0.58,0.58}
\definecolor{grey59}{rgb}{0.59,0.59,0.59}
\definecolor{grey5}{rgb}{0.05,0.05,0.05}
\definecolor{grey60}{rgb}{0.60,0.60,0.60}
\definecolor{grey61}{rgb}{0.61,0.61,0.61}
\definecolor{grey62}{rgb}{0.62,0.62,0.62}
\definecolor{grey63}{rgb}{0.63,0.63,0.63}
\definecolor{grey64}{rgb}{0.64,0.64,0.64}
\definecolor{grey65}{rgb}{0.65,0.65,0.65}
\definecolor{grey66}{rgb}{0.66,0.66,0.66}
\definecolor{grey67}{rgb}{0.67,0.67,0.67}
\definecolor{grey68}{rgb}{0.68,0.68,0.68}
\definecolor{grey69}{rgb}{0.69,0.69,0.69}
\definecolor{grey6}{rgb}{0.06,0.06,0.06}
\definecolor{grey70}{rgb}{0.70,0.70,0.70}
\definecolor{grey71}{rgb}{0.71,0.71,0.71}
\definecolor{grey72}{rgb}{0.72,0.72,0.72}
\definecolor{grey73}{rgb}{0.73,0.73,0.73}
\definecolor{grey74}{rgb}{0.74,0.74,0.74}
\definecolor{grey75}{rgb}{0.75,0.75,0.75}
\definecolor{grey76}{rgb}{0.76,0.76,0.76}
\definecolor{grey77}{rgb}{0.77,0.77,0.77}
\definecolor{grey78}{rgb}{0.78,0.78,0.78}
\definecolor{grey79}{rgb}{0.79,0.79,0.79}
\definecolor{grey7}{rgb}{0.07,0.07,0.07}
\definecolor{grey80}{rgb}{0.80,0.80,0.80}
\definecolor{grey81}{rgb}{0.81,0.81,0.81}
\definecolor{grey82}{rgb}{0.82,0.82,0.82}
\definecolor{grey83}{rgb}{0.83,0.83,0.83}
\definecolor{grey84}{rgb}{0.84,0.84,0.84}
\definecolor{grey85}{rgb}{0.85,0.85,0.85}
\definecolor{grey86}{rgb}{0.86,0.86,0.86}
\definecolor{grey87}{rgb}{0.87,0.87,0.87}
\definecolor{grey88}{rgb}{0.88,0.88,0.88}
\definecolor{grey89}{rgb}{0.89,0.89,0.89}
\definecolor{grey8}{rgb}{0.08,0.08,0.08}
\definecolor{grey90}{rgb}{0.90,0.90,0.90}
\definecolor{grey91}{rgb}{0.91,0.91,0.91}
\definecolor{grey92}{rgb}{0.92,0.92,0.92}
\definecolor{grey93}{rgb}{0.93,0.93,0.93}
\definecolor{grey94}{rgb}{0.94,0.94,0.94}
\definecolor{grey95}{rgb}{0.95,0.95,0.95}
\definecolor{grey96}{rgb}{0.96,0.96,0.96}
\definecolor{grey97}{rgb}{0.97,0.97,0.97}
\definecolor{grey98}{rgb}{0.98,0.98,0.98}
\definecolor{grey99}{rgb}{0.99,0.99,0.99}
\definecolor{grey9}{rgb}{0.09,0.09,0.09}
\definecolor{grey}{rgb}{0.75,0.75,0.75}
\definecolor{honeydew1}{rgb}{0.94,1.00,0.94}
\definecolor{honeydew2}{rgb}{0.88,0.93,0.88}
\definecolor{honeydew3}{rgb}{0.76,0.80,0.76}
\definecolor{honeydew4}{rgb}{0.51,0.55,0.51}
\definecolor{honeydew}{rgb}{0.94,1.00,0.94}
\definecolor{hotpink}{rgb}{1.00,0.41,0.71}
\definecolor{indianred}{rgb}{0.80,0.36,0.36}
\definecolor{ivory1}{rgb}{1.00,1.00,0.94}
\definecolor{ivory2}{rgb}{0.93,0.93,0.88}
\definecolor{ivory3}{rgb}{0.80,0.80,0.76}
\definecolor{ivory4}{rgb}{0.55,0.55,0.51}
\definecolor{ivory}{rgb}{1.00,1.00,0.94}
\definecolor{khaki1}{rgb}{1.00,0.96,0.56}
\definecolor{khaki2}{rgb}{0.93,0.90,0.52}
\definecolor{khaki3}{rgb}{0.80,0.78,0.45}
\definecolor{khaki4}{rgb}{0.55,0.53,0.31}
\definecolor{khaki}{rgb}{0.94,0.90,0.55}
\definecolor{lavenderblush}{rgb}{1.00,0.94,0.96}
\definecolor{lavender}{rgb}{0.90,0.90,0.98}
\definecolor{lawngreen}{rgb}{0.49,0.99,0.00}
\definecolor{lemonchiffon}{rgb}{1.00,0.98,0.80}
\definecolor{lightblue}{rgb}{0.68,0.85,0.90}
\definecolor{lightcoral}{rgb}{0.94,0.50,0.50}
\definecolor{lightcyan}{rgb}{0.88,1.00,1.00}
\definecolor{lightgoldenrod}{rgb}{0.93,0.87,0.51}
\definecolor{lightgoldenrod}{rgb}{0.98,0.98,0.82}
\definecolor{lightgray}{rgb}{0.83,0.83,0.83}
\definecolor{lightgreen}{rgb}{0.56,0.93,0.56}
\definecolor{lightgrey}{rgb}{0.83,0.83,0.83}
\definecolor{lightpink}{rgb}{1.00,0.71,0.76}
\definecolor{lightsalmon}{rgb}{1.00,0.63,0.48}
\definecolor{lightsea}{rgb}{0.13,0.70,0.67}
\definecolor{lightsky}{rgb}{0.53,0.81,0.98}
\definecolor{lightslate}{rgb}{0.47,0.53,0.60}
\definecolor{lightslate}{rgb}{0.47,0.53,0.60}
\definecolor{lightslate}{rgb}{0.52,0.44,1.00}
\definecolor{lightsteel}{rgb}{0.69,0.77,0.87}
\definecolor{lightyellow}{rgb}{1.00,1.00,0.88}
\definecolor{limegreen}{rgb}{0.20,0.80,0.20}
\definecolor{linen}{rgb}{0.98,0.94,0.90}
\definecolor{magenta1}{rgb}{1.00,0.00,1.00}
\definecolor{magenta2}{rgb}{0.93,0.00,0.93}
\definecolor{magenta3}{rgb}{0.80,0.00,0.80}
\definecolor{magenta4}{rgb}{0.55,0.00,0.55}
\definecolor{magenta}{rgb}{1.00,0.00,1.00}
\definecolor{maroon1}{rgb}{1.00,0.20,0.70}
\definecolor{maroon2}{rgb}{0.93,0.19,0.65}
\definecolor{maroon3}{rgb}{0.80,0.16,0.56}
\definecolor{maroon4}{rgb}{0.55,0.11,0.38}
\definecolor{maroon}{rgb}{0.69,0.19,0.38}
\definecolor{mediumaquamarine}{rgb}{0.40,0.80,0.67}
\definecolor{mediumblue}{rgb}{0.00,0.00,0.80}
\definecolor{mediumorchid}{rgb}{0.73,0.33,0.83}
\definecolor{mediumpurple}{rgb}{0.58,0.44,0.86}
\definecolor{mediumsea}{rgb}{0.24,0.70,0.44}
\definecolor{mediumslate}{rgb}{0.48,0.41,0.93}
\definecolor{mediumspring}{rgb}{0.00,0.98,0.60}
\definecolor{mediumturquoise}{rgb}{0.28,0.82,0.80}
\definecolor{mediumviolet}{rgb}{0.78,0.08,0.52}
\definecolor{midnightblue}{rgb}{0.10,0.10,0.44}
\definecolor{mintcream}{rgb}{0.96,1.00,0.98}
\definecolor{mistyrose}{rgb}{1.00,0.89,0.88}
\definecolor{moccasin}{rgb}{1.00,0.89,0.71}
\definecolor{navajowhite}{rgb}{1.00,0.87,0.68}
\definecolor{navyblue}{rgb}{0.00,0.00,0.50}
\definecolor{navy}{rgb}{0.00,0.00,0.50}
\definecolor{oldlace}{rgb}{0.99,0.96,0.90}
\definecolor{olivedrab}{rgb}{0.42,0.56,0.14}
\definecolor{orange1}{rgb}{1.00,0.65,0.00}
\definecolor{orange2}{rgb}{0.93,0.60,0.00}
\definecolor{orange3}{rgb}{0.80,0.52,0.00}
\definecolor{orange4}{rgb}{0.55,0.35,0.00}
\definecolor{orangered}{rgb}{1.00,0.27,0.00}
\definecolor{orange}{rgb}{1.00,0.65,0.00}
\definecolor{orchid1}{rgb}{1.00,0.51,0.98}
\definecolor{orchid2}{rgb}{0.93,0.48,0.91}
\definecolor{orchid3}{rgb}{0.80,0.41,0.79}
\definecolor{orchid4}{rgb}{0.55,0.28,0.54}
\definecolor{orchid}{rgb}{0.85,0.44,0.84}
\definecolor{palegoldenrod}{rgb}{0.93,0.91,0.67}
\definecolor{palegreen}{rgb}{0.60,0.98,0.60}
\definecolor{paleturquoise}{rgb}{0.69,0.93,0.93}
\definecolor{paleviolet}{rgb}{0.86,0.44,0.58}
\definecolor{papayawhip}{rgb}{1.00,0.94,0.84}
\definecolor{peachpuff}{rgb}{1.00,0.85,0.73}
\definecolor{peru}{rgb}{0.80,0.52,0.25}
\definecolor{pink1}{rgb}{1.00,0.71,0.77}
\definecolor{pink2}{rgb}{0.93,0.66,0.72}
\definecolor{pink3}{rgb}{0.80,0.57,0.62}
\definecolor{pink4}{rgb}{0.55,0.39,0.42}
\definecolor{pink}{rgb}{1.00,0.75,0.80}
\definecolor{plum1}{rgb}{1.00,0.73,1.00}
\definecolor{plum2}{rgb}{0.93,0.68,0.93}
\definecolor{plum3}{rgb}{0.80,0.59,0.80}
\definecolor{plum4}{rgb}{0.55,0.40,0.55}
\definecolor{plum}{rgb}{0.87,0.63,0.87}
\definecolor{powderblue}{rgb}{0.69,0.88,0.90}
\definecolor{purple1}{rgb}{0.61,0.19,1.00}
\definecolor{purple2}{rgb}{0.57,0.17,0.93}
\definecolor{purple3}{rgb}{0.49,0.15,0.80}
\definecolor{purple4}{rgb}{0.33,0.10,0.55}
\definecolor{purple}{rgb}{0.63,0.13,0.94}
\definecolor{red1}{rgb}{1.00,0.00,0.00}
\definecolor{red2}{rgb}{0.93,0.00,0.00}
\definecolor{red3}{rgb}{0.80,0.00,0.00}
\definecolor{red4}{rgb}{0.55,0.00,0.00}
\definecolor{red}{rgb}{1.00,0.00,0.00}
\definecolor{rosybrown}{rgb}{0.74,0.56,0.56}
\definecolor{royalblue}{rgb}{0.25,0.41,0.88}
\definecolor{saddlebrown}{rgb}{0.55,0.27,0.07}
\definecolor{salmon1}{rgb}{1.00,0.55,0.41}
\definecolor{salmon2}{rgb}{0.93,0.51,0.38}
\definecolor{salmon3}{rgb}{0.80,0.44,0.33}
\definecolor{salmon4}{rgb}{0.55,0.30,0.22}
\definecolor{salmon}{rgb}{0.98,0.50,0.45}
\definecolor{sandybrown}{rgb}{0.96,0.64,0.38}
\definecolor{seagreen}{rgb}{0.18,0.55,0.34}
\definecolor{seashell1}{rgb}{1.00,0.96,0.93}
\definecolor{seashell2}{rgb}{0.93,0.90,0.87}
\definecolor{seashell3}{rgb}{0.80,0.77,0.75}
\definecolor{seashell4}{rgb}{0.55,0.53,0.51}
\definecolor{seashell}{rgb}{1.00,0.96,0.93}
\definecolor{sienna1}{rgb}{1.00,0.51,0.28}
\definecolor{sienna2}{rgb}{0.93,0.47,0.26}
\definecolor{sienna3}{rgb}{0.80,0.41,0.22}
\definecolor{sienna4}{rgb}{0.55,0.28,0.15}
\definecolor{sienna}{rgb}{0.63,0.32,0.18}
\definecolor{skyblue}{rgb}{0.53,0.81,0.92}
\definecolor{slateblue}{rgb}{0.42,0.35,0.80}
\definecolor{slategray}{rgb}{0.44,0.50,0.56}
\definecolor{slategrey}{rgb}{0.44,0.50,0.56}
\definecolor{snow1}{rgb}{1.00,0.98,0.98}
\definecolor{snow2}{rgb}{0.93,0.91,0.91}
\definecolor{snow3}{rgb}{0.80,0.79,0.79}
\definecolor{snow4}{rgb}{0.55,0.54,0.54}
\definecolor{snow}{rgb}{1.00,0.98,0.98}
\definecolor{springgreen}{rgb}{0.00,1.00,0.50}
\definecolor{steelblue}{rgb}{0.27,0.51,0.71}
\definecolor{tan1}{rgb}{1.00,0.65,0.31}
\definecolor{tan2}{rgb}{0.93,0.60,0.29}
\definecolor{tan3}{rgb}{0.80,0.52,0.25}
\definecolor{tan4}{rgb}{0.55,0.35,0.17}
\definecolor{tan}{rgb}{0.82,0.71,0.55}
\definecolor{thistle1}{rgb}{1.00,0.88,1.00}
\definecolor{thistle2}{rgb}{0.93,0.82,0.93}
\definecolor{thistle3}{rgb}{0.80,0.71,0.80}
\definecolor{thistle4}{rgb}{0.55,0.48,0.55}
\definecolor{thistle}{rgb}{0.85,0.75,0.85}
\definecolor{tomato1}{rgb}{1.00,0.39,0.28}
\definecolor{tomato2}{rgb}{0.93,0.36,0.26}
\definecolor{tomato3}{rgb}{0.80,0.31,0.22}
\definecolor{tomato4}{rgb}{0.55,0.21,0.15}
\definecolor{tomato}{rgb}{1.00,0.39,0.28}
\definecolor{turquoise1}{rgb}{0.00,0.96,1.00}
\definecolor{turquoise2}{rgb}{0.00,0.90,0.93}
\definecolor{turquoise3}{rgb}{0.00,0.77,0.80}
\definecolor{turquoise4}{rgb}{0.00,0.53,0.55}
\definecolor{turquoise}{rgb}{0.25,0.88,0.82}
\definecolor{violetred}{rgb}{0.82,0.13,0.56}
\definecolor{violet}{rgb}{0.93,0.51,0.93}
\definecolor{wheat1}{rgb}{1.00,0.91,0.73}
\definecolor{wheat2}{rgb}{0.93,0.85,0.68}
\definecolor{wheat3}{rgb}{0.80,0.73,0.59}
\definecolor{wheat4}{rgb}{0.55,0.49,0.40}
\definecolor{wheat}{rgb}{0.96,0.87,0.70}
\definecolor{whitesmoke}{rgb}{0.96,0.96,0.96}
\definecolor{white}{rgb}{1.00,1.00,1.00}
\definecolor{yellow1}{rgb}{1.00,1.00,0.00}
\definecolor{yellow2}{rgb}{0.93,0.93,0.00}
\definecolor{yellow3}{rgb}{0.80,0.80,0.00}
\definecolor{yellow4}{rgb}{0.55,0.55,0.00}
\definecolor{yellowgreen}{rgb}{0.60,0.80,0.20}
\definecolor{yellow}{rgb}{1.00,1.00,0.00}

\def\fsu5{$\cal{F}$-$SU(5)$}
\def\bfsu5{$\boldsymbol{\mathcal{F}}$-$\boldsymbol{SU(5)}$}

\def\m1half{$M_{1/2}$}
\def\m3half{$M_{3/2}$}
\def\m32{$M_{32}$}

\begin{document}

\title{A Multi-Axis Best Fit to the Collider Supersymmetry Search:
	\\The Aroma of Stops and Gluinos at the $\mathbf{\sqrt{s} = 7}$~TeV LHC}

\author{Tianjun Li}

\affiliation{State Key Laboratory of Theoretical Physics and Kavli Institute for Theoretical Physics China (KITPC),
Institute of Theoretical Physics, Chinese Academy of Sciences, Beijing 100190, P. R. China}

\affiliation{George P. and Cynthia W. Mitchell Institute for Fundamental Physics and Astronomy,
Texas A$\&$M University, College Station, TX 77843, USA}

\author{James A. Maxin}

\affiliation{George P. and Cynthia W. Mitchell Institute for Fundamental Physics and Astronomy,
Texas A$\&$M University, College Station, TX 77843, USA}

\author{Dimitri V. Nanopoulos}

\affiliation{George P. and Cynthia W. Mitchell Institute for Fundamental Physics and Astronomy,
Texas A$\&$M University, College Station, TX 77843, USA}

\affiliation{Astroparticle Physics Group, Houston Advanced Research Center (HARC),
Mitchell Campus, Woodlands, TX 77381, USA}

\affiliation{Academy of Athens, Division of Natural Sciences,
28 Panepistimiou Avenue, Athens 10679, Greece}

\author{Joel W. Walker}

\affiliation{Department of Physics, Sam Houston State University,
Huntsville, TX 77341, USA}


\begin{abstract}

In {\it Profumo di SUSY}, we presented evidence that the CMS and ATLAS Collaborations
may have already registered a handful of deftly camouflaged supersymmetry events at the LHC
in the multijet channels.  Here, we explore the prospect for corroboration of this suggestion
from five additional  CMS and ATLAS search strategies targeting the production of light stops and gluinos
at lower jet counts, which variously depend on heavy flavor tagging and the inclusion or
exclusion of associated leptons.  The current operating phase of the $\mathbf{\sqrt{s} = 7}$~TeV LHC is highly
conducive to the production of gluinos and light stops, given the supersymmetric particle mass hierarchy
$M({\widetilde{t_1}}) < M({\widetilde{g}}) < M({\widetilde{q}})$ that naturally evolves from the
dynamics of the model named No-Scale \fsu5 that we presently study.  Moreover, some tension
persists against the Standard Model data-driven and Monte-Carlo generated background predictions
in certain LHC searches of this variety.  We demonstrate that the 1-$\sigma$ overlap of the allowed
supersymmetric event production for these seven search methodologies roundly envelops the most
favorable phenomenological subspace of \fsu5, whilst handily generating a 125~GeV Higgs boson mass.
In order to test the statistical significance of any correlations across the simulated \fsu5 collider
response in these seven search strategies, we implement a multi-axis $\chi^2$ fitting procedure, yielding
a best overall match in the vicinity of $M_{1/2} = 610$~GeV, corresponding to light stop and gluino
masses of approximately 665~GeV and 830~GeV.  Consequently, we suggest that No-Scale \fsu5 is a better global fit to the studied LHC
data than the SM alone, and moreover that its predictions appear to be meaningfully correlated with observed
low-statistics excesses across a wide variety of specialized search strategies.  We suspect that the already
collected $5~{\rm fb}^{-1}$ of integrated luminosity will be sufficient to either condense or disperse 
the delicate aroma of stops and gluinos that suffuses the early search.

\end{abstract}



\preprint{ACT-04-12, MIFPA-12-09}

\maketitle


\section{Introduction}

The Large Hadron Collider (LHC) at CERN has to date delivered an integrated luminosity of up to 5~${\rm fb}^{-1}$
of proton-proton collisions at a center-of-mass beam energy of $\mathbf{\sqrt{s} = 7}$~TeV,
with a further 15 ${\rm fb}^{-1}$ anticipated during 2012, in tandem with an upgrade to $\mathbf{\sqrt{s} = 8}$~TeV.
The CMS and ATLAS Collaborations have analyzed the first 1--2 ${\rm fb}^{-1}$ of data in their
search for signals of supersymmetry (SUSY), thus far revealing no significant excesses beyond the
Standard Model (SM) expectations. The absence of any definitive signal has imposed severe constraints
onto the viable parameter spaces of the baseline supersymmetric models, leading pessimists to question whether
there is even a SUSY framework extant in nature to discover at all. Nevertheless, a few select search
methodologies have exhibited curious indications of strain between the maximum reasonable expectation for the
background and the number of observations in the
data~\cite{PAS-SUS-11-003,ATLAS-CONF-2011-130,Aad:2011ib,Aad:2011qa,PAS-SUS-11-019}.  This
begs the question of whether such tensions are mere statistical fluctuations, the inevitable
``look elsewhere'' styled distribution tails that become probable {\it somewhere} within an exhaustive search,
or perhaps rather early warning indicators of something much more significant yet to come.  Given any reasonable likelihood
that we {\it may} indeed be witnessing the commencement, in its earliest nascent phase, of a legitimate SUSY signal
effervescing to the surface, then it becomes imperative to closely examine these particular search strategies in unison
to uncover clues as to whether they might originate from a common underlying physics.  Such a multi-axis study could imply
a key connection not apparent from the individual searches in isolation, allowing for the directed {\it correlation}
across independent selection spaces to statistically distinguish a signal from the noise.  The demonstration
of a specifically detailed supersymmetric model that can expose tendencies for high SUSY visibility in the
appropriate searches, without damaging event overproduction in those cases where the observations are in stricter 
accord with the expected backgrounds, would suggestively hint that the currently observed stresses might
represent an authentic physical correlation after all.

In a previous work entitled {\it Profumo di SUSY}~\cite{Li:2011av}, we investigated a pair of early single 
inverse femtobarn LHC reports from CMS~\cite{PAS-SUS-11-003} and ATLAS~\cite{Aad:2011qa}, where data
observations for events with ultra-high jet multiplicities, minimally seven to nine, could be readily
extracted. In the time frame preceding release of the CMS and ATLAS reports~\cite{PAS-SUS-11-003,Aad:2011qa},
we had been strongly advocating for close scrutiny of precisely these sort of
events~\cite{Li:2011hr,Maxin:2011hy,Li:2011gh,Li:2011rp}, following a realization that the unique
spectrum of our favored SUSY construction, and in particular its rather light stop squark $\widetilde{t}_1$
and gluino $\widetilde{g}$, would lead to a strong and distinctive signal in these particular channels. 
Consequently, we were gratified that an initial inquiry into such events could be undertaken at the LHC,
and buoyed by the observation of scant, yet nonetheless tantalizing, excesses in events with $\ge$7--9
jets.  We thus undertook a carefully detailed study in Ref.~\cite{Li:2011av}.  During the intermission between those
early CMS and ATLAS reports~\cite{PAS-SUS-11-003,Aad:2011qa} and the present time frame, numerous parallel
high luminosity studies from CMS and ATLAS have appeared to compliment those discriminated by jet count.  In
the present work, we thus broaden our horizon to encompass the wider panorama of search strategies squarely directed 
at the detection of pair-produced gluinos and light stops, including light stops transpiring from gluino mediated
decays.  This is a calculated augmentation, given that our model is auspiciously predisposed to the production of gluinos
and light stops in the present running phase of the LHC.  Moreover, it facilitates precisely the sort of
cross-correlation between independent experimental signatures previously described.  This will be quantified in
the present work by application of a $\chi^2$ statistical test in seven degrees of freedom.  If the prior fleeting
scent~\cite{Li:2011av} of supersymmetry in the multijet data was corporeal, then that same aroma should linger
also in certain of the wider stop and gluino searches; if it was a wishfully conjured mirage, then even the memory
must fade in the cleansing light of new data.


\section{The No-Scale $\mathbf{\cal{F}}$-$\mathbf{SU(5)}$ Model}

The context of our study on the correlation of light squark and gluino SUSY searches is a model named No-Scale
\fsu5~\cite{Li:2010ws, Li:2010mi,Li:2010uu,Li:2011dw, Li:2011hr, Maxin:2011hy, Li:2011xu,Li:2011in,Li:2011gh,Li:2011rp,Li:2011fu,Li:2011xg,Li:2011ex,Li:2011av,Li:2011ab}.
No-Scale \fsu5 is defined by the convergence of the ${\cal F}$-lipped $SU(5)$~\cite{Barr:1981qv,Derendinger:1983aj,Antoniadis:1987dx}
grand unified theory (GUT), two pairs of hypothetical TeV scale vector-like supersymmetric multiplets (dubbed {\it flippons}) of mass $M_{\rm V}$
with origins in local ${\cal F}$-theory~\cite{Jiang:2006hf,Jiang:2009zza,Jiang:2009za,Li:2010dp,Li:2010rz} model building, and the dynamically
established boundary conditions of No-Scale Supergravity~\cite{Cremmer:1983bf,Ellis:1983sf, Ellis:1983ei, Ellis:1984bm, Lahanas:1986uc}.
This construction inherits all of the most beneficial phenomenology of the flipped $SU(5) \times U(1)_{\rm X}$~\cite{Nanopoulos:2002qk,Barr:1981qv,Derendinger:1983aj,Antoniadis:1987dx}
gauge group structure, as well as all of the valuable theoretical motivation of No-Scale Supergravity~\cite{Cremmer:1983bf,Ellis:1983sf, Ellis:1983ei, Ellis:1984bm, Lahanas:1986uc}.
A substantially more detailed theoretical treatment of the model under analysis is available in the cited references,
including a rather thorough summary in the Appendix of Ref.~\cite{Maxin:2011hy}.

Since mass degenerate superpartners for the known SM fields are not observed,
SUSY must itself be broken around the TeV scale. In the Constrained Minimal
Supersymmetric Standard Model (CMSSM) and minimal supergravities
(mSUGRA)~\cite{Chamseddine:1982jx}, this occurs first in a hidden sector, and the secondary
propagation by gravitational interactions into the observable sector is parameterized by universal
SUSY-breaking ``soft terms'' which include the gaugino mass $M_{1/2}$, scalar mass $M_0$ and the
trilinear coupling $A$. The ratio of the low energy Higgs vacuum expectation values (VEVs) $\tan \beta$,
and the sign of the SUSY-preserving Higgs bilinear mass term $\mu$ are also undetermined, while the
magnitude of the $\mu$ term and its bilinear soft term $B_{\mu}$ are determined by the $Z$-boson mass $M_Z$
and $\tan \beta$ after electroweak symmetry breaking (EWSB). In the simplest No-Scale scenario,
$M_0$=A=$B_{\mu}$=0 at the unification boundary, while the complete collection of low energy SUSY
breaking soft-terms evolve down with a single non-zero parameter $M_{1/2}$. Consequently, the particle
spectrum will be proportional to $M_{1/2}$ at leading order, rendering the bulk ``internal'' physical
properties invariant under an overall rescaling. The matching condition between the low-energy value of
$B_\mu$ that is demanded by EWSB and the high-energy $B_\mu = 0$ boundary is notoriously difficult to
reconcile under the renormalization group equation (RGE) running. The present solution relies on
modifications to the $\beta$-function coefficients that are generated by radiative loops containing
the vector-like {\it flippon} multiplets.  By coupling to the Higgs boson, the {\it flippons} will moreover 
have an impact on the Higgs boson mass $m_h$~\cite{Moroi:1992zk,Babu:2008ge,Huo:2011zt}, resulting in a 3-4~GeV
upward shift in $m_h$, handily generating a Higgs mass of 124-126~GeV~\cite{Li:2011ab} that is in fine
accord with the recent ATLAS and CMS reports~\cite{Collaboration:2012tx,Collaboration:2012si}.

Pertinent to the present work, we have previously demonstrated the range of the \fsu5 model
space that is adherent to a set of firm ``bare-minimal'' phenomenological constraints~\cite{Li:2011xu},
including consistency with the world average top-quark mass $m_{\rm t}$~\cite{:1900yx},
the dynamically established boundary conditions of No-Scale supergravity, radiative electroweak
symmetry breaking, the centrally observed WMAP7 CDM relic density~\cite{Komatsu:2010fb}, and precision LEP
constraints on the lightest CP-even Higgs boson $m_{h}$~\cite{Barate:2003sz,Yao:2006px} and other light
SUSY chargino and neutralino mass content.  We have moreover established a highly constrained
subspace, dubbed the {\it Golden Strip}~\cite{Li:2010mi,Li:2011xu,Li:2011xg}, that is noteworthy for
its capacity to additionally conform to the phenomenological limits on rare processes that are established
by measurement of the muon anomalous magnetic moment $(g_{\mu}-2)/2$ and the branching ratios of the
flavor-changing neutral current decays $b \to s\gamma$ and $B_S^0 \rightarrow \mu^+\mu^-$.  A similarly
favorable {\it Silver Strip} slightly relaxes the constraints imposed by $(g_{\mu}-2)$.


\section{\fsu5 Stops and Gluinos}

Production of the light stop $\widetilde{t}_1$ and gluino $\widetilde{g}$ at the early LHC in the first
5~${\rm fb^{-1}}$ of integrated luminosity and naturally accounting for a 125~GeV Higgs boson tend to be mutually exclusive
goals for the traditional MSSM constructions. In particular, the mechanism for elevation of the Higgs mass will
typically correspond to squark and gluino masses which are far too heavy to have yet crept above the SM
background for the initial $\sqrt{s} = 7$~TeV operating phase of the LHC. For instance, it has been
suggested that in the CMSSM and mSUGRA, the only viable remnant of solution space that has thus far
survived the rapidly encroaching LHC constraints while delivering a 125~GeV Higgs boson mass requires a super-heavy scalar
mass $m_0$ of 10--20 TeV~\cite{Baer:2012uy}, which is well beyond reach of the LHC operating energy, both currently and
into the future.  The additional contributions from the vector-like {\it flippons} are the key
to differentiating \fsu5 from the CMSSM and mSUGRA. The {\it flippon} loops allow a 125~GeV Higgs boson in
conjunction with a light TeV-scale SUSY spectrum. In contrast, the CMSSM and mSUGRA require very large values
of the trilinear A-term, which pushes $m_0$ to very large values, such that the extremely massive stop and sbottom
squark masses lead to severe electroweak fine-tuning~\cite{Baer:2012uy}. No-Scale \fsu5 takes
advantage of the same strongness of the Higgs to top quark coupling that provides the primary lifting of the
SUSY Higgs mass to generate a hierarchically light partner stop in the SUSY mass-splitting.  However, this
rather generic mechanism is not in itself enough. The model further leverages the same vector-like
multiplets which provide the secondary Higgs mass perturbation to flatten the RGE running of universal
color-charged gaugino mass $M_3$, blocking the standard logarithmic enhancement of the gluino mass at
low energies, and producing the distinctive mass ordering
$M({\widetilde{t_1}}) < M({\widetilde{g}}) < M({\widetilde{q}})$ of a light stop and gluino, both
substantially lighter than all other squarks. The stability of this distinctive mass hierarchy is
manifest across the entire model space, a hierarchy that is not
precisely replicated in any MSSM constructions of which we are aware.

Indeed, it is specifically because the light stop $\widetilde{t}_1$ and gluino $\widetilde{g}$ are less
massive than the heavier bottom squarks $\widetilde{b}_1$ and $\widetilde{b}_2$ and the first and second generation
left and right squarks $\widetilde{q}_R$ and $\widetilde{q}_L$ that we are afforded a uniquely
distinctive test signature for \fsu5 at the LHC.  This spectrum generates a characteristic event
topology starting from the pair production of heavy squarks $\widetilde{q}$ and/or gluinos
$\widetilde{g}$ in the initial hard scattering process, with each squark likely to yield a quark-gluino
pair $\widetilde{q} \rightarrow q \widetilde{g}$ in the cascade decay. The gluino will tend
to decay via QCD to a typical 2-jet final state as $\widetilde{g} \rightarrow q \overline{q} \widetilde{\chi}_1^0$,
though at an atypically low 60\% branching ratio. The weakly interacting
lightest neutralino $\widetilde{\chi}_1^0$ escapes the detector unseen, leaving only an imprint of
missing energy.  This leaves allowance for the production of light stops through gluino decays $\widetilde{g} \rightarrow \widetilde{t}_1 \overline{t}$
at a relatively high rate of 40\%, where the light stops decay as $\widetilde{t}_1 \rightarrow t \widetilde{\chi}_1^0$
at 58\% and as $\widetilde{t}_1 \rightarrow b \widetilde{\chi}_1^{\pm}$ at 32\%.  We note that the
intermediate light stop may tend to be off shell, particularly for the lighter \fsu5 spectra,
below a gaugino mass $M_{1/2}$ of about 700~GeV.

The repercussions of these final states are two-fold.  Firstly, it is expected that each gluino will produce
{\it at least} four hard jets 40\% of the time. Processes such as this may then consistently exhibit a net
product of eight or more hard jets emergent from a single squark-squark, squark-gluino, or gluino-gluino
pair production event. When the further process of jet fragmentation is allowed after the primary hard
scattering events and the sequential cascade decay chain, this will ultimately result in a
spectacular signal of ultra-high multiplicity final state events. Events with very high multiplicity
jets have received little study in legacy experiments such as LEP and those at the Tevatron, though
fortunately such analyses are now beginning to receive more than just sporadic attention at LHC.
Recognizing the prospect of a conveniently encoded SUSY signal within such multijet events, we
optimistically anticipate a near-term expansion search horizon in the high multiplicity regime.

A second impact of the \fsu5 final states could be discovery of the light stop and gluino production itself.
Considering the high production rate of gluino mediated light stops at 40\%, those SUSY searches
currently focused more intently on the $\widetilde{g} \rightarrow \widetilde{t}_1 t$,
$\widetilde{t}_1 \rightarrow b \widetilde{\chi}_1^{\pm}$, and $\widetilde{t}_1 \rightarrow t \widetilde{\chi}_1^0$
channels may also expect to reap tangible benefits within this construction.  In contrast
to the relatively unexplored region of ultra-high jet multiplicities, searches more directly focused on
gluino and light stop production are gaining maturity. These searches
typically concentrate on final product states of b-jets (heavy flavor tagging) and leptons, along with smaller multiplicities
of jets. Thus, we would not be surprised at all if an initial conclusive signal discovery emanated from
these search methodologies. In fact, a dual signal emergent in gluino and stop production search
strategies and in the ultra-high jet multiplicity events will be highly suggestive of \fsu5 origins.

A further consequence of the accessible mass of the \fsu5 light stop in the present operational phase of the
LHC is the more pronounced direct production cross-section of the light stops from the hard scattering
event. An inspection of the direct production cross-sections for squarks, gluinos, and light stops,
along with the branching ratios, yields an expectation that about 15-20\% of light stop production at the LHC
in an \fsu5 framework would be pair-production directly from the hard scattering collision. This is in
contrast to other MSSM based constructions, where it is not uncommon for less than 1\% of all light stops to be produced directly.


\section{The LHC SUSY Search\label{sct:search}}

We now focus on seven ongoing LHC SUSY search strategies, of which five are substantially orthogonal in construction,
that are sensitive to the \fsu5 final states comprised of stops and gluinos decaying into some quantity of jets.
Each one of these seven event selection methodologies exhibits at least slight positive strain against the expectation
for the SM background, a correlation that we shall demonstrate may not be coincidental.  First, we offer a concise summary
of each search strategy, then present the \fsu5 contribution to each in Section~\ref{sct:correlations}, followed by a multi-axis
$\chi^2$ best fit against the full contingent of selection strategies in Section~\ref{sct:chi2}.

\subsection{CMS Purely Hadronic Large Jet Multiplicities}

This search is detailed in Ref.~\cite{PAS-SUS-11-003} and based upon a data sample of 1.1~${\rm fb^{-1}}$.
All hadronic events with high $p_T$ are discriminated by jet count, allowing for smooth
extrapolation of events with very high jet multiplicities. The primary cuts are $H_T \ge$375~GeV,
$E_T^{Miss} \ge$100~GeV, and $p_T >$50~GeV. We use the data sample with no $\alpha_T$ cut, which we have
argued is actively biased against events with high multiplicities of jets~\cite{Maxin:2011hy}. We apply a further
cut on jet count, retaining only those events with greater than or equal to nine jets. This search strategy
is very favorable for exposing an \fsu5 signal emanating from the sequential cascade decays of gluinos,
squarks, and light stops to many jets. We have studied this search methodology in some detail in Ref.~\cite{Li:2011av},
and we now reprise that analysis with the intent of revealing any potentially hidden correlations with a much
more broad sampling of contemporary LHC SUSY searches.

\subsection{ATLAS Large Jet Multiplicities}

The fine points for this search can be found in Ref.~\cite{Aad:2011qa}, in a study based upon
1.34~${\rm fb^{-1}}$ of data. Here again, all events are segregated by jet count, permitting a straightforward extraction
of events of all high multiplicity jet counts. Additionally, four key combinations of the jet count and transverse momentum $p_T$
per jet thresholds are isolated in the tables for detailed study.  We choose to keep only those events with at least 7 jets with
$p_T >$80~GeV for the case of $E_T^{Miss}/\sqrt{H_T} >$3.5.  As with the preceding SUSY search, this scenario will also
be sensitive to the large \fsu5 multijet final states. Likewise, we invested much detail in the
analysis of this search strategy in Ref.~\cite{Li:2011av}, which we again carry over in the interest of
exposing correlations within a more comprehensive range of possible channels for a SUSY discovery.
The present incarnation of this search does differ with our prior report in one regard: we have opted
in this work to employ the cone jet finding algorithm provided with {\tt PGS4}~\cite{PGS4} rather
than the $k_t$ jet alternative.

\subsection{ATLAS B-jets plus Lepton}

The first undertaking of this ATLAS search strategy is defined in Ref.~\cite{ATLAS-CONF-2011-130}, employing a
data sample of 1.03 ${\rm fb^{-1}}$. The requirement here is at least four jets, a minimum of one b-jet,
precisely one lepton, $p_T >$50~GeV for all jets, $E_T^{Miss} >$80~GeV, and $M_{eff} >$600~GeV. In our
analysis here, we choose to harness the data-driven background findings. This strategy will be sensitive
to large cross-sections of $\widetilde{g}\widetilde{g}$ production with large branching ratios for
$\widetilde{g} \rightarrow \widetilde{t}_1 t$, as is expected in \fsu5. Therefore, this strategy is
very sensitive to gluino-mediated light stop production in \fsu5, which is currently in a very favorable production
phase at the LHC.  However, we must note that the simplified model interpretation in Ref.~\cite{ATLAS-CONF-2011-130}
assumes a 100\% branching ratio for $\widetilde{t}_1 \rightarrow b \widetilde{\chi}_1^{\pm}$,
whereas the \fsu5 branching ratio is only 32\%. Thus, we caution that the \fsu5 spectra may be
misrepresented in the generic imposition of model limits.

\subsection{ATLAS B-jets plus Lepton SR1-D}

The ATLAS B-jets plus Lepton search in the previous subsection has been updated to an extent in
Ref.~\cite{ATLAS-CONF-2012-003} for 2.05 ${\rm fb^{-1}}$.  While still implementing the same
pre-selection cuts as Ref.~\cite{ATLAS-CONF-2011-130}, the cuts on the leading jet $p_T$ and $M_{eff}$
have been altered, in turn significantly affecting the surviving background sample, and possibly any
embedded signal as well. Therefore, we consider the shift in the final results to be consequential enough
to warrant an independent analysis of Refs.~\cite{ATLAS-CONF-2011-130}
and~\cite{ATLAS-CONF-2012-003}.  In Ref.~\cite{ATLAS-CONF-2012-003}, the two strategies of
interest here are the SR1-D and SR1-E. In the case of SR1-D, the search parameters have been updated such
that the $p_T$ for the leading jet has been raised to $p_T>$60~GeV and the cut on effective mass has been
increased to $M_{eff} >$ 700~GeV.

\subsection{ATLAS B-jets plus Lepton SR1-E}

The additional SR1-E scenario of Ref.~\cite{ATLAS-CONF-2011-130}, over and above the further cuts
implemented in SR1-D, has elevated the missing energy component to $E_T^{miss} >$ 200~GeV. In our analysis
to follow in Section~\ref{sct:correlations}, we shall clearly discriminate between the three ATLAS B-jets plus Lepton cases
due to the substantial impact that the retuned cuts have on the data sample.  Each scenario will therefore
illustrate a unique state of the SUSY discovery program.

\subsection{ATLAS Purely Hadronic Events}

This study is based upon the search methodology in Ref.~\cite{Aad:2011ib}, using a data sample of 1.04~${\rm fb^{-1}}$.
We apply the ``High Mass'' cuts, consisting of at least four jets, $p_T >$80~GeV ($p_T >$130~GeV
for the leading jet), no lepton, $E_T^{Miss} >$130~GeV, and $M_{eff} >$1100~GeV. The intent of the ``High
Mass'' strategy is to extend a maximal reach into the SUSY mass spectrum. Sensitivity will be high for
models with large cross-sections of pair-produced $\widetilde{g}\widetilde{g}$,
$\widetilde{g}\widetilde{q}$, and $\widetilde{q}\widetilde{q}$, where $\widetilde{q} \rightarrow q \widetilde{\chi}_1^0$
and $\widetilde{g} \rightarrow q \overline{q} \widetilde{\chi}_1^0$. As indicated
earlier, the gluinos in \fsu5 are lighter than all squarks except
the light stop, hence the $\widetilde{q} \rightarrow q \widetilde{g}$ channel will prevail more than 90\% of the
time for the $\widetilde{q}_R$ and two-thirds of the time for the $\widetilde{q}_L$. Thus, the
$\widetilde{q} \rightarrow q \widetilde{\chi}_1^0$ path is comparatively suppressed in \fsu5.  However,
with the rate of gluino to jets at 60\% for $\widetilde{g} \rightarrow q \overline{q} \widetilde{\chi}_1^0$, this
search should remain sensitive to the \fsu5 gluino production.  Since leptons are explicitly excluded, the QCD
background here is expected to be troublesome.  The implementation of the ``High Mass'' cuts certainly
alleviate the QCD predicament to some degree, although the signal may also be diminished in scope as well.

\subsection{CMS Jet-Z Balance}

We employ the search of Ref.~\cite{PAS-SUS-11-019} here, which is based upon a data sample of 2.1~${\rm fb^{-1}}$.
The JZB method concentrates on states containing a Z-boson, jets and missing energy. The
advantage here is that the contribution from $Z \rightarrow l^+l^-$ is clean, and the Z+jets contribution
can be predicted. This is sensitive to SUSY $\widetilde{g}\widetilde{g}$ production, with the gluino
decay to a neutralino via $\widetilde{g} \rightarrow q \overline{q} \widetilde{\chi}_2^0$, followed
by $\widetilde{\chi}_2^0 \rightarrow Z \widetilde{\chi}_1^0$. However, the branching ratio of
$\widetilde{g} \rightarrow q \overline{q} \widetilde{\chi}_2^0$ is only 18\% in \fsu5, while the
$\widetilde{\chi}_2^0 \rightarrow Z \widetilde{\chi}_1^0$ is a mere 0.34\%. Thus, with only a 0.06\%
probability of a $\widetilde{g} \rightarrow q \overline{q} Z \widetilde{\chi}_1^0$ transition,
expectations are that this channel will experience high suppression in \fsu5, and hence provide no
observable SUSY signals within the 2.1 ${\rm fb^{-1}}$ data sample.


\section{Simulation and Error Analysis\label{sct:simulation}}

\begin{figure*}[htp]
        \centering
        \includegraphics[width=0.70\textwidth]{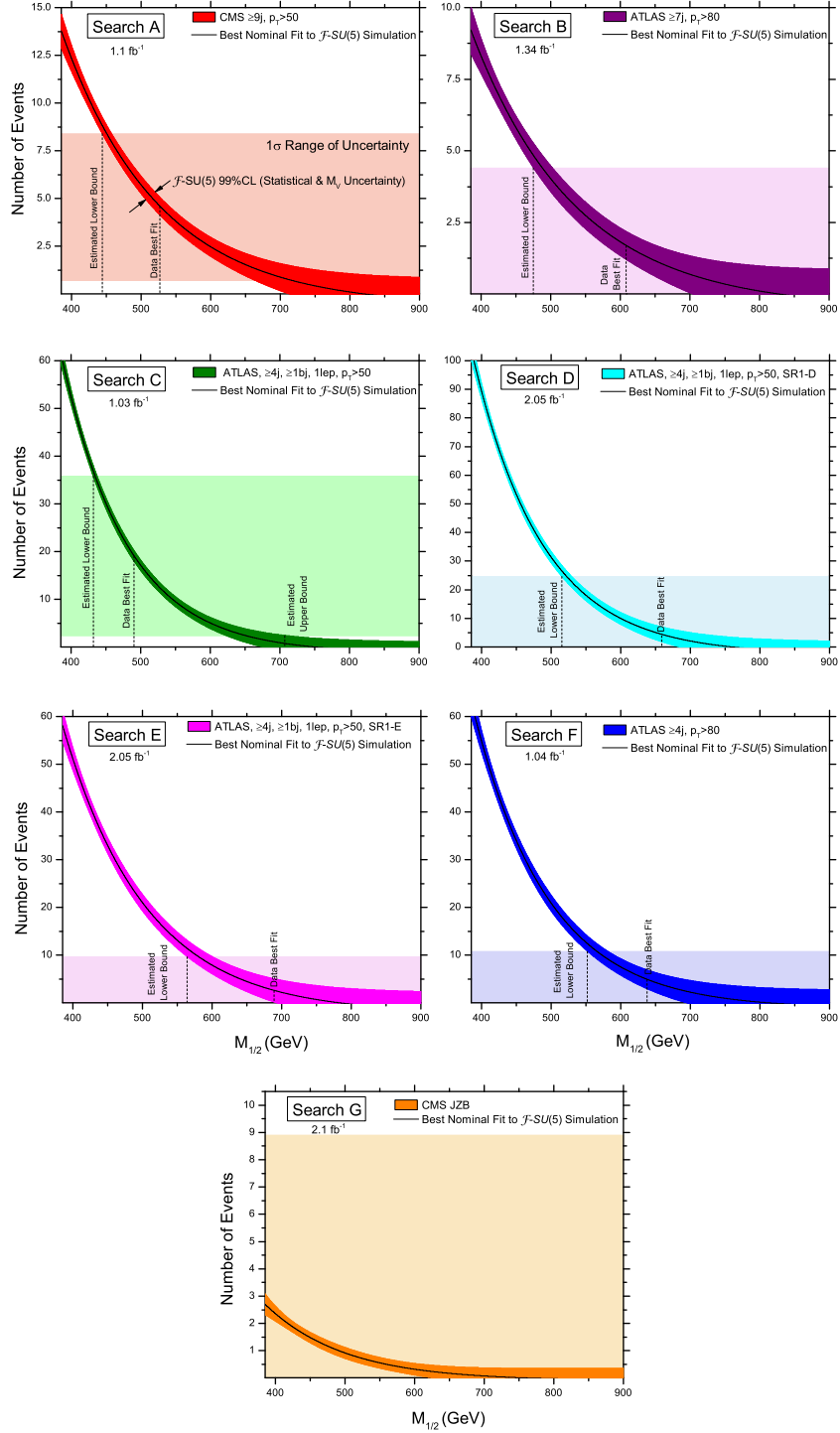}
        \caption{Event counts for \fsu5 are plotted as a function of the gaugino mass $M_{1/2}$. The span of $M_{1/2}$
	in each plot space consists of the minimum $M_{1/2}$ = 385~GeV and maximum $M_{1/2}$ = 900~GeV allowed
	by application of the bare-minimal phenomenological constraints of Ref.~\cite{Li:2011xu}. The thickness of
	each curve is the consequence of a superposition of statistical uncertainty and the flexible range on the
	{\it flippon} mass $M_V$, where variance of $M_V$ has a minor effect on the event counts. The median line transversing
	the thickness is the best nominal fit to the \fsu5 event count data. The rectangular shaded regions identify
	the maximum and minimum number of events allowed to maintain consistency with the CMS and ATLAS reported
	SM background and data observations. Therefore, the intersection of the \fsu5 curves with the rectangular
	range of uncertainty isolates the estimated upper and lower boundaries on $M_{1/2}$ in \fsu5 that preserve
	uniformity with the CMS and ATLAS results for each individual search.}
        \label{fig:7plex}
\end{figure*}

The explicit event selection scenarios from each of the seven CMS and ATLAS search strategies A--G
discussed in the previous section are applied to a representative sampling of the viable
\fsu5 parameter space satisfying the bare-minimal constraints of Ref.~\cite{Li:2011xu}. The
resulting event counts are extrapolated to the full phenomenologically viable model space for each of the
seven cases, as depicted in Fig.~\ref{fig:7plex}.  To achieve this result, we employ a detailed Monte Carlo
collider-detector simulation of all 2-body SUSY processes based on the standard
{\tt MadGraph}~\cite{Stelzer:1994ta,MGME} suite, including the {\tt MadEvent}~\cite{Alwall:2007st},
{\tt PYTHIA}~\cite{Sjostrand:2006za} and {\tt PGS4}~\cite{PGS4} chain.  We employ the ATLAS and CMS
detector specification cards provided with {\tt PGS4}, and specify the cone jet clustering algorithm in all cases,
with an angular scale parameter $\Delta R$ of 0.5 for CMS, and 0.4 for ATLAS.  The results are
filtered according to a careful replication of the individual SUSY search selection cuts,
using a script {\tt CutLHCO} of our own design~\cite{cutlhco}. SUSY particle mass calculations were performed using
{\tt MicrOMEGAs 2.1}~\cite{Belanger:2008sj}, employing a proprietary modification of
the {\tt SuSpect 2.34}~\cite{Djouadi:2002ze} codebase to run the {\it flippon}-enhanced RGEs.

An estimated uncertainty on the fitting between the simulated \fsu5 event counts and the mass scale $M_{1/2}$ is
computed, representing a 99\% confidence level. This narrow region of uncertainty on
the \fsu5 simulations, represented by the band width in Fig.~\ref{fig:7plex}, characterizes a combination of statistical
uncertainty and variations in the vector-like {\it flippon} mass due to a yet unknown resolution of this
$M_V$ parameter.  Although the {\it flippon} mass can in
principle be limited to a more constrained range in order to facilitate a 124-126~GeV Higgs boson mass, we have
depicted in Fig.~\ref{fig:7plex} a range of uncertainty that would be inclusive of all {\it flippon}
masses, for the sake of completeness. For reference, the nominal best $M_{1/2}$ fit to each individual study is
further included. We overlay the upper and lower boundaries of uncertainty on the CMS and ATLAS derived background
and data observations, displayed as rectangular shaded regions in Fig.~\ref{fig:7plex}. This allows us
to clearly demonstrate those regions of the \fsu5 model space that comply with each individual search
methodology, noting estimated lower bounds on $M_{1/2}$, and in one particular case, also an upper bound on $M_{1/2}$.

We have attempted to normalize the treatment of error propagation across the
various studies under consideration. In all seven cases, we are provided an
uncertainty on the SM background estimate by the collaboration. In one of
these cases~\cite{PAS-SUS-11-003}, the uncertainty is carefully extracted bin-by-bin
from the graphical presentation and summed in quadrature, while the remaining
six reports provide direct numerical values. Whenever statistical and systematic
errors are provided separately, we likewise combine these in quadrature. When
given a choice between data-driven and Monte Carlo analyses, we favor the background
estimate and associated (reduced) error provided by the data-driven methodology.
If differential upper and lower bounds are provided, we adopt the larger of the
error statistics. The central background estimates and associated 1-$\sigma$
deviations are listed in the second column of Table~\ref{tab:signals} in Section~\ref{sct:correlations}
for each study.

In general, the collaboration studies have not provided an explicit uncertainty on
the observed data counts, but one may confidently expect Poisson statistics to apply
here, and an error scaling that goes like the square root of the recorded events.
Specifically, we adopt a factor of $\sqrt{(1+{\rm Data})}$ across the board, which compares
satisfactorily with a graphical extraction of the event frequency bounds published in
Ref.~\cite{PAS-SUS-11-003}. The relevant observations are summarized in the third column
of Table~\ref{tab:signals}. Since the nominal target for SUSY event contributions is
the observed excess of the recorded data over the SM background estimate, this statistical
uncertainty on the event count must be combined in quadrature with the previously
described uncertainty of the background estimate itself. Reassuringly, a doubling
of this statistic to the 2-$\sigma$ level generates a very favorable match to the
95\% confidence outer bounds on SUSY counts over observed excesses that are reported in
Refs.\cite{Aad:2011qa,ATLAS-CONF-2012-003}.

Finally, any reported excess must be compared against our Monte Carlo simulation of
the \fsu5 model space. The statistical errors on our procedure are very small, being
minimized by substantial oversampling of the integrated luminosity. Moreover, the
application of a combined full-model space fit of the expected event count against the
gaugino mass $M_{1/2}$ further smoothes out statistical variations. There remains some
uncertainty due to variation of the vector-like {\it flippon} mass scale $M_{\rm V}$,
but this higher order effect is nicely accounted for by the demonstrated width of the curves
in Fig.~\ref{fig:7plex}, and is thus not further considered outside that context. On the other hand, the
systematic errors on our procedure may be rather large, and are quite difficult to reliably
estimate in a systematic manner. We have opted to employ a factor of
$\sqrt{(1+{\rm Observed~Excess})}$, where the observed excess that the experiments report 
over the expected backgrounds are tallied in the fifth column of Table~\ref{tab:signals}.
Although this quantity is actually more naturally suited to describe the sources of statistical
error than the systematic, it may still be a reasonable estimate for the systematic error, in
the absence of other concrete options. Since the comparison of our simulation against the
experimental results constitutes an implicit second level of subtraction, the corresponding error
must again be combined in quadrature with the net experimental error on the data excess.
It is this final statistic that is employed to tally the minimum and maximum permissible event
count variation at the 1-$\sigma$ level in columns 4 and 6 of Table~\ref{tab:signals}. It is
also used as the denominator of our subsequent multi-axis $\chi^2$ best fit against the
seven LHC SUSY search strategies that have been outlined.


\section{The \fsu5 Correlations\label{sct:correlations}}

Notable in Fig.~\ref{fig:7plex} is the consistency which the \fsu5 model enjoys with all seven search
schemes, with each generally sharing a similar locally favored region of the parameter space. Excluding the highly
suppressed CMS JZB search, the smallest lower bound is given by the ATLAS bjet and lepton search, at a little less
than $M_{1/2} = 440$~GeV, with the ATLAS and CMS multijet searches also
posting sub-500~GeV gaugino masses. In close proximity, the ATLAS bjet and lepton SR1-D search does
however set a lower bound on $M_{1/2}$ just slightly above 500~GeV. As forecasted, the minuscule
production of $q \overline{q} Z \widetilde{\chi}_1^0$ does in fact mightily subdue the number of \fsu5
observations in the CMS JZB cutting technique, such that no lower boundary can be ascribed to $M_{1/2}$
using the JZB tactic. Thus, it is interesting that five of the seven searches fix the lower bound on
$M_{1/2}$ at about 500~GeV or less. The residual two probes, namely the ATLAS purely hadronic ``High Mass'' cuts and
ATLAS bjet and lepton SR1-E, call for a lower limit on $M_{1/2}$ in the neighborhood of 550~GeV.
Consequently, we can conclude that the \fsu5 model space just above about $M_{1/2} = 550$~GeV is alive and
well after application of all CMS and ATLAS 1-2 ${\rm fb^{-1}}$ constraints, with the model space above
$M_{1/2} = 500$~GeV perfectly tolerated in five of the seven searches. Linking these $M_{1/2}$ model
parameter values to experimentally vital scales, $M_{1/2} = 550$~GeV corresponds to a light stop
$\widetilde{t}_1$ mass of about 600~GeV and a gluino $\widetilde{g}$ of about 750~GeV; $M_{1/2} = 500$~GeV
correlates to a light stop of about 540~GeV and a gluino of around 690~GeV.

\begin{figure*}[htp]
        \centering
        \includegraphics[width=0.80\textwidth]{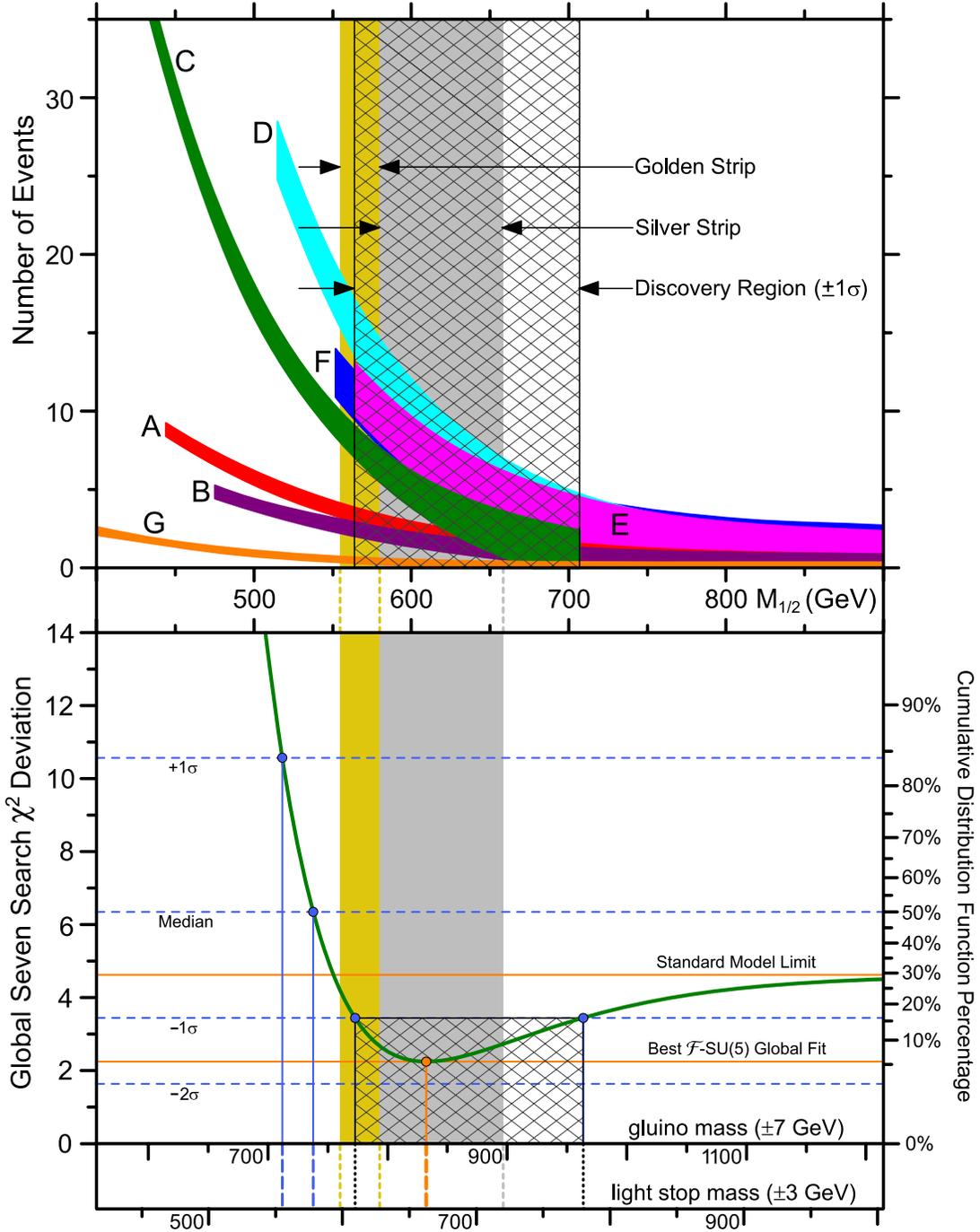}
        \caption{In the upper panel, we superimpose event counts for the seven search methodologies studied in this work,
	labeling the 1-$\sigma$ overlap between these strategies as the ``Discovery Region''.
	The lower panel depicts a multi-axis $\chi^2$ fit to the same set of seven search strategies, where the
	cumulative distribution function percentage demarcated on the right-hand axis dips to a minimum of 5.5\%
	for the best overall fit at $M_{1/2} = 610$~GeV.  The range of the parameter space that provides a
	better fit than the median at 1-$\sigma$ significance is in broad agreement with the previously noted
	Discovery Region, and both notably intersect with the phenomenologically favored {\it Golden Strip} and
	{\it Silver Strip} regions.  The light stop and gluino masses (in~GeV) corresponding to each $M_{1/2}$ value
	have been inserted onto the lower horizontal axes.  Uncertainties of a few GeV exist in the mapping of
	the axis labels for both cases, due to higher order fluctuations arising from variation in the top quark
	mass $m_t$ and {\it flippon} mass $M_V$.}
        \label{fig:discovery}
\end{figure*}

Six of the seven schemes require no upper bound on $M_{1/2}$ as a result of there existing no inordinate
number of excess events. Nonetheless, the ATLAS bjet and lepton study does exhibit an excess even at the
maximum 1-$\sigma$ Standard Model limit, when applying the data-driven background statistics. The experimental
collaborations at LHC are striving for data-driven backgrounds in their SUSY searches, hence we believe
the choice of data-driven over Monte-Carlo generated to be justified. If this small residue is indeed substantive,
then we are compelled to enforce an upper boundary on $M_{1/2}$ at about 710~GeV, which
corresponds to a 785~GeV light stop mass and 970~GeV gluino mass. The interesting material
result is the constitution of a narrow strip between 565 $\lesssim M_{1/2} \lesssim$ 710~GeV corresponding
to an overlapping ``Discovery Region'' that is favorable for a potential finding of SUSY at ATLAS and CMS.
In the upper panel of Fig.~\ref{fig:discovery}, we visually demonstrate this overlap by resketching those segments of
the curves from Fig.~\ref{fig:7plex} that maintain compatibility at 1-$\sigma$ with experimental uncertainties on each search.  We
remark that the upper limit imposed here is somewhat provisional, pending a more substantial 
accumulation of collision data.  In particular, the collaborations themselves maintain that the SM alone remains
essentially compatible with the data, to an acceptable statistical significance.  Nevertheless, the current single standard
deviation limits taken at face value suggest an upper boundary for substantiation or exclusion on the \fsu5 model space at
about $M_{1/2} \simeq$~710~GeV ($\widetilde{t}_1 \simeq$~785~GeV and $\widetilde{g} \simeq$~970~GeV).

The spatial synchronicity displayed by the Discovery Region in Fig.~\ref{fig:discovery} with the
phenomenologically derived {\it Golden Strip} at $555 \le M_{1/2} \le 580$~GeV
and {\it Silver Strip} at $580 \le M_{1/2} \le$~658~GeV is rather striking,
and embodies the recurring weight of strong correlation between ostensibly independent experimental data
points that we have become increasingly accustomed to observing throughout our extended study 
of the No-Scale \fsu5 model.  It is an essential prerequisite that any high-energy framework of nature
discovered by precise measurements at LHC must correctly simultaneously account for the WMAP-7 measured
relic density, the top quark mass and other precision electroweak parameters, and the rare-process constraints.
This is a test that No-Scale \fsu5 seems well poised to pass.

The SM background, data observations, and uncertainty statistics are detailed in Table~\ref{tab:signals}.
The ``Total SM'' and ``Data'' tabulations are those reported by CMS and ATLAS.  The
``${\rm Signal_{Min}}$'' and ``${\rm Signal_{Max}}$'' entries describe the 1-$\sigma$ confines on the range of
excess SUSY events, as shown in Fig.~\ref{fig:7plex}.  We further display in this table nine markers of the expected
event count in terms of $M_{1/2}$ to numerically illustrate the relevant progression through the region of interest.
Those entries highlighted in green depict consistency with the allowed range of uncertainty. For the ranges
displayed in Table~\ref{tab:signals}, those points in the vicinity of $M_{1/2} =$~600--650~GeV are best capable of
simultaneously satisfying all seven LHC search strategies.   Note that the width of the Discovery Region in
Fig.~\ref{fig:discovery} is somewhat wider than this boundary, due to the thickness of the fitting to event counts.
Five benchmark spectra are listed in Table~\ref{tab:masses} by their model parameters, along with those supersymmetric
particle masses directly relevant to our discussion.  The presented benchmarks are chosen specifically to highlight
the nominal best fits to various SUSY search strategies from Section~\ref{sct:search}.
Specifically note the Higgs boson masses near 125~GeV for each, in tandem with the light stop and gluino.

\begin{table*}[htpb]
\caption{A comparison of the \fsu5 event counts for each of the seven search strategies discussed in
	Section~\ref{sct:search}. The ``Total SM'' and ``Data'' columns display the reported SM background and
	data observations by the CMS and ATLAS Collaborations, with the ``Observed Excess'' column being the
	difference between the nominal values of the SM and data. The ``${\rm Signal_{Min}}$'' and ``${\rm Signal_{Max}}$''
	values are the minimum and maximum allowed event counts in order to be consistent with the background and observations,
	where the error analysis determination is elaborated in Section~\ref{sct:simulation}.
	The ``$M_{1/2}$'' columns (in~GeV) contain the \fsu5 event counts for each of the $M_{1/2}$ given, for a mass
	range chosen to represent those regions of the parameter space consistent with the LHC searches under study. The
	highlighted event counts are those that conform to the stated range within the ``${\rm Signal_{Min}}$''
	and ``${\rm Signal_{Max}}$'' upper and lower boundaries.}
{\centering
\footnotesize
\begin{tabular}{|c||c|c||c|c|c||c|c|c|c|c|c|c|c|c|c|}\cline{7-15}
\multicolumn{6}{c|}{} & \multicolumn{9}{c|}{\rm $M_{1/2}$}\\ \cline{1-15}
$\rm Search$&$\rm~~ Total~SM ~~$&$\rm Data$&$\rm Signal_{Min}$&$\rm Observed~ Excess$&$\rm Signal_{Max}$&$~~475~~$&$~~500~~$&$~~525~~$&$~~550~~$&$~~575~~$&$~~600~~$&$~~625~~$&$~~650~~$&$~~675~~$ \\ \hline	\hline
$\rm A$&$3.4 \pm 0.7$&$8$&$0.7$&$4.6$&$8.4$&${\color{green4} 6.9}$&${\color{green4} 5.7}$&${\color{green4} 4.7}$&${\color{green4} 3.8}$&${\color{green4} 3.1}$&${\color{green4} 2.5}$&${\color{green4} 2.0}$&${\color{green4} 1.5}$&${\color{green4} 1.2}$ \\ \hline
$\rm B$&$1.3 \pm 0.9$&$3$&$0.0$&$1.7$&$4.4$&$4.8$&${\color{green4} 4.0} $&${\color{green4} 3.3}$&${\color{green4} 2.7} $&$ {\color{green4} 2.2}$&${\color{green4} 1.8}$&${\color{green4} 1.5}$&${\color{green4} 1.2}$&${\color{green4} 0.9}$ \\ \hline
$\rm C$&$54.9 \pm13.6$&$74$&$2.4$&$19.1$&$35.8$&${\color{green4} 22.7} $&${\color{green4} 17.1} $&${\color{green4} 12.8} $&${\color{green4} 9.5} $&$ {\color{green4} 7.0}$&${\color{green4} 5.0}$&${\color{green4} 3.6}$&${\color{green4} 2.4}$&$1.6$ \\ \hline
$\rm D$&$77.0 \pm 18.4$&$81$&$0.0$&$4.0$&$24.7$&$40.9 $&$31.3 $&${\color{green4} 23.8} $&${\color{green4} 18.0} $&$ {\color{green4} 13.5}$&${\color{green4} 10.0}$&${\color{green4} 7.2}$&${\color{green4} 5.1}$&${\color{green4} 3.4}$ \\ \hline
$\rm E$&$14.4 \pm 5.4$&$17$&$0.0$&$2.6$&$9.7$&$26.3  $&$ 21.0 $&$ 16.6 $&$ 13.1 $&$ 10.2 $&${\color{green4} 7.9}$&$ {\color{green4} 6.0}$&${\color{green4} 4.5}$&${\color{green4} 3.2}$ \\ \hline
$\rm F$&$13.1 \pm 3.1$&$18$&$0.0$&$4.9$&$10.8$&$26.6$&$20.9 $&$16.3$&$12.7 $&$ {\color{green4}9.8}$&${\color{green4}7.5}$&${\color{green4}5.7}$&${\color{green4} 4.3}$&${\color{green4} 3.1}$ \\ \hline
$\rm G$&$7.0 \pm 2.6$&$11$&$0.0$&$ 4.0$&$8.9$&${\color{green4} 1.2} $&${\color{green4} 0.9} $&${\color{green4} 0.7} $&$ {\color{green4} 0.6}$&${\color{green4} 0.4} $&${\color{green4} 0.3} $&$ {\color{green4} 0.2}$&${\color{green4} 0.2}$&${\color{green4} 0.1}$ \\ \hline
\end{tabular}}
\label{tab:signals}
\end{table*}

\begin{table*}[htpb]
\caption{Higgs boson and sparticle masses (in~GeV) are given for five benchmark gaugino masses $M_{1/2}$,
	representative of a best fit for each search methodology examined in this work. The light stop
	$\widetilde{t}_1$ and gluino $\widetilde{g}$ columns have been highlighted to reflect their discovery
	mass ranges, all of which should be accessible at the $\sqrt{s} = 8$~TeV LHC in 2012, whereas the lighter
	points may have already been substantively probed by the $\sqrt{s} = 7$~TeV LHC during the 2011 run. The $M_{1/2} = 518$~GeV point
	is the representative benchmark of Ref.~\cite{Li:2011ab}. The $M_{1/2}$ = 610~GeV point is also
	indicative of the precise minimum of the multi-axis $\chi^2$ fit described in Section~\ref{sct:chi2}.
	Significantly each of the five benchmarks cataloged here can moreover handily generate a 124-126~GeV Higgs mass.}
\centering
\footnotesize
\begin{tabular}{|c|c|c|c|c||c|c|c|c|c|c|c|c|c|c|c|c|c|}\hline
$\rm Search$&$\rm M_{1/2}$&$\rm M_V$&$\rm m_t$&$\rm tan\beta$&$\rm \widetilde{\chi}_1^0$&$\rm Higgs$&$\rm \widetilde{\chi}_2^0$&$\rm \widetilde{\chi}_1^{\pm}$&$\rm \widetilde{t}_1$&$\rm \widetilde{g}$&$\rm \widetilde{b}_1$&$\rm \widetilde{t}_2$&$\rm \widetilde{b}_2$&$\rm \widetilde{u}_R$&$\rm \widetilde{d}_R$&$\rm \widetilde{u}_L$&$\rm \widetilde{d}_L$ \\ \hline	\hline
$\rm A$&$ 518$&$ 1640$&$ 174.4$&$ 20.65$&$99 $&$ \textbf{125.4}$&$216 $&$216 $&${\color {blue} 558} $&${\color {blue} 704} $&$934 $&$982 $&$1046 $&$1053 $&$1094 $&$1144 $&$1147 $ \\ \hline
$\rm B$&$ 610$&$ 2500 $&$ 174.3 $&$21.44  $&$ 121 $&$ \textbf{124.4}$&$260  $&$260  $&${\color {blue} 669 } $&${\color {blue} 826 }$&$1076 $&$ 1117 $&$1194 $&$ 1207$&$ 1252 $&$ 1312$&$ 1314 $ \\ \hline
$\rm C$&$ ~485~$&$~1475~ $&$~174.3~ $&$ ~20.40~$&$~92~ $&$~ \textbf{125.5}~ $&$~200~ $&$ ~200~$&$ ~{\color {blue} 518}~$&$~{\color {blue} 661}~ $&$~881 ~$&$~932~ $&$ ~989~$&$ ~994~$&$~1034~ $&$~1080 ~$&$~1083~ $ \\ \hline
$\rm D,E$&$ 675$&$2950  $&$ 174.4 $&$ 21.87 $&$136  $&$\textbf{124.6} $&$291  $&$291  $&${\color {blue} 746 } $&${\color {blue} 910 } $&$1179  $&$1215 $&$ 1301 $&$1318  $&$ 1367 $&$ 1433 $&$ 1435 $ \\ \hline
$\rm F$&$ 638$&$ 2505$&$174.4 $&$ 21.63$&$ 127$&$ \textbf{124.9}$&$ 273$&$ 273$&$ {\color {blue} 703}$&${\color {blue} 861} $&$1123 $&$1161 $&$1243 $&$1258 $&$1305 $&$1367 $&$1369 $ \\ \hline
\end{tabular}
\label{tab:masses}
\end{table*}


\section{A Multi-Axis $\chi^2$ Fit\label{sct:chi2}}

We have implemented a $\chi^2$ test in order to establish the optimal correspondence
between the No-Scale \fsu5 model space and the ongoing LHC SUSY search, and also to
gauge the overall statistical significance of the resulting best fit. To facilitate
this task, it was necessary to first establish a continuous functional relationship
between the gaugino mass $M_{1/2}$ and the expected event count for each SUSY search
strategy under consideration. It should be noted that this process is greatly simplified,
and the result thereby lent much greater parsimony, by the fact that the spectrum is
generated at leading order by only the single mass parameter. To proceed, we generously
sampled the \fsu5 model space at nineteen representative benchmark combinations of
$(M_{1/2},M_{\rm V},m_{\rm t}~{\rm and}~\tan\beta)$, generating a detailed Monte Carlo
collider-detector simulation, including the careful application of relevant selection cuts,
as described in Section~\ref{sct:simulation}.
For each search strategy, a satisfactory continuous empirical fit was obtained
by linear regression at quadratic order to the log-log distribution of event counts
vs. $M_{1/2}$.

The $\chi^2$ test statistic, which is expected to asymptotically approach the
formal $\chi^2_N$ distribution, is defined as
\begin{equation}
\chi^2 (M_{1/2}) = \sum_{i=1}^N \left\{ \frac{{\left({\rm Events}(M_{1/2})_i - {\rm Excess}_i \right)}^2}{\sigma_i^2} \right\} \,
\end{equation}
where ${\rm Events}(M_{1/2})_i$ is the continuous fit to the number of SUSY events expected
from \fsu5 at the given mass scale, ${\rm Excess}_i$ is the observed excess from Table~\ref{tab:signals}
and $\sigma_i^2$ is the square of the single standard deviation error described in Section~\ref{sct:correlations},
all under the $i^{th}$ set of selection cuts, with $N = 7$.
The resulting function is plotted in the lower panel of Fig.~\ref{fig:discovery}, demonstrating a distinct
minimum in the vicinity of $M_{1/2} = 610$~GeV, corresponding to light stop and gluino masses of approximately
665~GeV and 830~GeV. The significance of the fit is established by comparison
with the formal $\chi^2_N$ probability distribution with $N$ degrees of freedom, which establishes
the likelihood of a given value for the $\chi^2$ test statistic under application of the ``null hypothesis'',
{\it i.e.}~where all signal deviations from the observed excess are attributed to uncorrelated Gaussian
fluctuations about the $\sigma_i$.

The figure of merit is the cumulative distribution function (CDF) of $\chi^2_N$,
which indicates the fraction of randomized trials under action of the null hypothesis that should be expected to
produce a lower value (better fit) of the $\chi^2$ test statistic than some given threshold value. The median
value of the CDF for $N=7$ is $6.35$, and the double-sided $\pm 1,2 \sigma$ CDF values, corresponding to the traditional
Gaussian percentage thresholds of 2.28\%, 15.87\%, 84.13\% and 97.73\% (centrally encapsulating 68\% and 95\%),
fall at $\chi^2 = (1.64, 3.44, 10.57~{\rm and}~16.27)$, respectively. The best fit value of $M_{1/2}$ produces
a rather small $\chi^2$ of 2.24, corresponding to a CDF of 5.5\%, immediately on the cusp of the range generally
considered to represent a statistically significant deviation from the null hypothesis. The fit produced in the
SM limit, the asymptote of the soft high-mass boundary in the lower panel of Fig.~\ref{fig:discovery}, is also
reasonably satisfactory, with $\chi^2 = 4.62$, and a CDF of 29.4\%. Models disfavored for overproduction at
the 2- and 1-$\sigma$ limits have mass scales $M_{1/2}$ of 501~GeV and 518~GeV, respectively, while the median fit
occurs at 538~GeV. Models favored by a negative deviation of one standard deviation or more in the CDF exist within the
range from $M_{1/2} = 564$~GeV to $M_{1/2} = 709$~GeV, with the best fit, again, around $M_{1/2} = 610$~GeV.


\section{The Once and Future LHC}

We close our analysis with a brief glance in postscript toward the future $\sqrt{s}$ = 8 TeV beam energy and 15 ${\rm fb^{-1}}$
of data expected in 2012, as well as the already collected 5 ${\rm fb^{-1}}$ at $\sqrt{s} = 7$~TeV that remains to be reported.
If the results presented in our study in fact represent persistent correlations in the data, and not merely statistical fluctuations,
then evidence of their verity should continue to ripen.  The ``Observed Excess'' in Table~\ref{tab:signals} exemplifies the slender
corridor with which we are attempting to extricate a signal of supersymmetry's presence in the data. All excesses but one
are less then five events, making precise extrapolations tenuous.  Moreover, the subtraction of large numbers (the net event count
and the expected SM background) to yield a small differential implies rather large proportional uncertainties, pushing the statistical
machinery to extremes.  This is a key reason that the SM asymptote maintains a reasonably favorably $\chi^2$ value in
Fig.~\ref{fig:discovery}.

Although (or because) we remain in a fledgling phase of the LHC data collection mission, and the status of plausible signal candidates 
is still tentative, this is also a period of incredibly rapid and dynamic development at the high energy and high intensity frontiers.
Firstly, the fact that the beam quality is still being tuned means that the time integrated luminosity continues to grow much more
rapidly than a linear trend.  Secondly, since the doubling interval for data collection is still somewhat low, newly accumulated statistics
make extremely strong fractional contributions to the combined knowledge over reasonable time scales.  A growth of the excess to a
minimum of ten events will assist in sharpening the analysis to a degree, and we suspect that a full analysis of the 2012 statistics
should put us well on the way toward a conclusive resolution to the matter.  Thirdly, the shortly anticipated upgrade to a $\sqrt{s} = 8$~TeV
beam will substantially enhance the expected SUSY event cross-sections; our Monte Carlo simulations attribute an improved
time efficiency in the collection of productive data on the order of two to this upgrade. 

Table~\ref{tab:610} reports the extrapolated signal significance at $5~{\rm fb}^{-1}$ for $\sqrt{s} = 7$~TeV,
computed as the ratio $S/\sqrt{B+1}$ of signal events $S$ to the square root of one plus the expected background $B$,
for each of the seven considered SUSY searches A--G, assuming viability of the central $\chi^2$ fit at $M_{1/2} = 610$~GeV. 
The ``discovery index'' (DI) calculates the required scaling on luminosity, reported in inverse femtobarns, that would be required
to establish a nominal signal significance of five. Also shown is the signal significance at $15~{\rm fb}^{-1}$ for $\sqrt{s} = 8$~TeV, summed with the statistics for $5~{\rm fb}^{-1}$ for $\sqrt{s} = 7$~TeV, for a total of $20~{\rm fb}^{-1}$. We see that four of the seven searches exceed the gold standard of 5.0 for signal significance for the total collected luminosity of $20~{\rm fb}^{-1}$ expected by the close of 2012.

\begin{table*}[htb]
    \caption{The extrapolated signal significance $S/\sqrt{B+1}$ at $5~{\rm fb}^{-1}$ for $\sqrt{s} = 7$~TeV is presented for
	each of the seven considered SUSY searches A--G, for $M_{1/2} = 610$~GeV. The ``discovery index'' (DI)
	calculates the required scaling on luminosity, reported in inverse femtobarns, that would be required to
	establish a nominal signal significance of five. Also shown is the signal significance at $15~{\rm fb}^{-1}$ for $\sqrt{s} = 8$~TeV, summed with the statistics for $5~{\rm fb}^{-1}$ for $\sqrt{s} = 7$~TeV, for a total of $20~{\rm fb}^{-1}$.}
{\centering
\footnotesize
\begin{tabular}{|c||c|c|c|c||c|c||c|}\cline{2-8}

\multicolumn{1}{c|}{} & \multicolumn{3}{c|}{$5~{\rm fb^{-1}}~@~7~{\rm TeV}$} & \multicolumn{1}{c||}{$7~{\rm TeV}$} & \multicolumn{2}{c||}{$15~{\rm fb^{-1}}~@~8~{\rm TeV}$} & \multicolumn{1}{c|}{$5~{\rm fb^{-1}}~@~7~{\rm TeV}~+~15~{\rm fb^{-1}}~@~8~{\rm TeV}$}\\ \cline{1-8}
$\rm ~~Search~~$&$~~{\cal F}-SU(5)~~$&$\rm ~~SM~~$&$\rm ~~S/ \sqrt{B+1}~~$&$\rm ~~DI~({\rm fb^{-1}})~~$&$~~{\cal F}-SU(5)~~$&$\rm ~~SM~~$&$\rm ~~S/ \sqrt{B+1}~~$ \\ \hline \hline

$\rm A$&$ 8.2$&$ 15.5$&$ 2.0$&$ 29~ $&$57$&$107$&$5.8$ \\ \hline
$\rm B$&$ 4.5$&$ 4.9 $&$ 1.9 $&$31~   $&$31$&$34$&$5.7$ \\ \hline
$\rm C$&$ 15.5$&$267 $&$0.9 $&$ 139~ $&$107$&$1842$&$2.7$ \\ \hline
$\rm D$&$ 16.3$&$188  $&$ 1.2 $&$ 88~  $&$113$&$1297$&$3.3$ \\ \hline
$\rm E$&$ 12.9$&$ 35$&$2.2 $&$ 26~ $&$89$&$242$&$6.1$ \\ \hline
$\rm F$&$ 24.0$&$ 63$&$3.0 $&$ 13.8~ $&$166$&$435$&$8.5$ \\ \hline
$\rm G$&$ 0.17$&$ 16.0$&$0.1 $&$ 69,204~ $&$1.2$&$110$&$0.1$ \\ \hline
\end{tabular}}
\label{tab:610}
\end{table*}


\section{Conclusions}

No conclusive indication of supersymmetry (SUSY) has been observed at the early LHC as the accumulation of
data advances toward 5~${\rm fb^{-1}}$, yet could enticing clues be germinating in the massive collection
of observations? This is the challenging question that presently attracts our interest, and which we may currently
attempt to address only by leaning upon the limited 1--2~${\rm fb^{-1}}$ of integrated luminosity amassed thus far.
Although we cannot argue for incontrovertible evidence of SUSY peeping beyond the Standard Model veil, we can
safely suggest that No-Scale \fsu5 is a better global fit to the data than the SM alone, and moreover that
its predictions appear to be meaningfully correlated with observed low-statistics excesses across a wide variety
of specialized search strategies, while gracefully avoiding devastating overproduction where events are not observed.
This is a strong statement in an era when the portion of phenomenologically viable MSSM and mSUGRA constructions
is diminishing rapidly, choked out by inconsistency with the Higgs measurements and advancing squark and gluino exclusion limits.

The No-Scale \fsu5 model, by virtue of its distinctive supersymmetric mass hierarchy of
$M({\widetilde{t_1}}) < M({\widetilde{g}}) < M({\widetilde{q}})$, possesses the signature event
fingerprint of a very high multiplicity of hadronic jets.  Moreover, the light stop and gluino masses,
possibly within reach of data collected during the $\sqrt{s} = 7$~TeV LHC run, renders
presently maturing searches aimed at stops and gluinos quite pertinent to the testing of \fsu5 as well.
Building upon a prior analysis of two existing searches conducted by CMS and ATLAS, we studied
five additional CMS and ATLAS search strategies, variously employing cuts on jets,
b-jets and leptons, designed to reveal light stop, gluino, and squark production. A number of
interesting conclusions were established.

Notably, we found that the entire region of the model
space for $M_{1/2} \gtrsim$ 440~GeV ($m_{\widetilde{t}_1} \gtrsim$ 460~GeV and
$m_{\widetilde{g}} \gtrsim$ 600~GeV) thrives in at least one of the seven search methodologies.
When all seven searches are combined, a lower bound of $M_{1/2} \gtrsim$ 565~GeV ($m_{\widetilde{t}_1} \gtrsim$ 615~GeV and
$m_{\widetilde{g}} \gtrsim$ 770~GeV) can be tentatively set.
Also intriguing is the potential for one experiment, the ATLAS search requiring at least
one b-jet and exactly one lepton, to demand an {\it upper} bound of $M_{1/2} \lesssim$ 710~GeV
($m_{\widetilde{t}_1} \lesssim$ 785~GeV and $m_{\widetilde{g}} \lesssim$ 970~GeV).  However, it should be
mentioned that more recent ATLAS results (also included in the present study) on similar, though not identical,
selection cuts do alleviate the need to mandate an upper bound on $M_{1/2}$.  The values of $M_{1/2}$ which are compatible
with all search strategies under present consideration in the 1-$\sigma$ overlap exist within the 
range of 565 $\lesssim M_{1/2} \lesssim$ 710~GeV, which we refer to as the Discovery Region.
In order to test the statistical significance of any correlations across the simulated \fsu5 collider
response in these seven search strategies, we implemented a multi-axis $\chi^2$ fitting procedure.  The best
overall match was obtained in the vicinity of $M_{1/2} = 610$~GeV (corresponding to light stop and gluino
masses of approximately 665~GeV and 830~GeV), where the $\chi^2_N$ cumulative distribution
function in seven parameters was reduced to 5.5\%.  The range of masses having a better fit than the median
at 1-$\sigma$ significance is 564--709~GeV, which is in excellent agreement with the simpler overlap statistic
just reported.

Both mechanisms produce a conspicuous overlap with the highly phenomenologically favorable
{\it Golden Strip} and {\it Silver Strip}, which add good agreement with rare process constraints on flavor
changing neutral currents and the anomalous magnetic moment of the muon to the broader capacity of the model
for respecting the WMAP7 relic density, the world average top-quark mass, radiative electroweak symmetry breaking,
precision LEP Higgs and SUSY constraints, and the dynamically established boundary conditions of No-Scale supergravity.
No less propitious is the ability to handily generate a 125~GeV Higgs boson
mass through additional loop process contributions from the vector-like {\it flippon} multiplets, producing
a fine accord with the statistical excesses recently reported by CMS and ATLAS in the mass range of 124-126~GeV.

The damage exacted onto the supersymmetric model landscape by the swiftly progressing LHC constraints
has been severe.  The tension between the growing likelihood of a 125~GeV Higgs boson mass, developing CMS
and ATLAS exclusion zones, and a supersymmetric spectrum light enough to be within reach of the current
operational phase of the LHC has greatly altered the conventional wisdom as to how a discovery
of supersymmetry would manifest at the LHC.  While the validation prospects for almost all prospective
models has greatly withered, the outlook for \fsu5 appears to have in fact brightened; in stark contrast,
it is perfectly capable of simultaneously striking each of these three targets.  A rare feat nowadays, indeed.


\begin{acknowledgments}
This research was supported in part
by the DOE grant DE-FG03-95-Er-40917 (TL and DVN),
by the Natural Science Foundation of China
under grant numbers 10821504 and 11075194 (TL),
by the Mitchell-Heep Chair in High Energy Physics (JAM),
and by the Sam Houston State University
2011 Enhancement Research Grant program (JWW).
We also thank Sam Houston State University
for providing high performance computing resources.
\end{acknowledgments}


\bibliography{bibliography}

\end{document}